\documentclass[reqno,a4paper,11pt]{article}
\pdfoutput=1
\usepackage{xcolor}

\usepackage{graphicx}
\usepackage[textwidth = 430 pt, textheight = 630 pt]{geometry}

\definecolor{MyDarkBlue}{rgb}{0.15,0.25,0.45}
\usepackage{epsfig,rotating}
\usepackage{amsmath,amssymb}
\usepackage{amsfonts}
\usepackage{mathrsfs}
\usepackage{bbm}
\usepackage[normalem]{ulem}
\usepackage{booktabs}

\usepackage{latexsym}
\usepackage{amsthm}
\usepackage[all,knot]{xy}
\xyoption{arc}

\usepackage[utf8x]{inputenc}

\usepackage{hyperref}
\hypersetup{
hypertexnames=false,
colorlinks=true,
citecolor=MyDarkBlue,
linkcolor=MyDarkBlue,
urlcolor=MyDarkBlue,
pdfauthor={Christian S\"amann and Lennart Schmidt},
pdftitle={Towards an M5-Brane Model I: A 6d Superconformal Field Theory},
pdfsubject={hep-th math-ph},
breaklinks=true
}

\usepackage{tikz}
\usepackage{mathtools}
\usepackage[all,knot]{xy}
\xyoption{arc}

\let\fn\footnote
\renewcommand{\footnote}[1]{\linespread{1.1}\fn{#1}\linespread{1.29}}

\makeatletter\renewcommand{\section}{\@startsection
{section}{1}{\z@}{-3.5ex plus -1ex minus
    -.2ex}{2.3ex plus .2ex}{\bf }}
\makeatletter\renewcommand{\subsection}{\@startsection{subsection}{2}{\z@}{-3.25ex
plus -1ex minus
   -.2ex}{1.5ex plus .2ex}{\bf }}
\makeatletter\renewcommand{\subsubsection}{\@startsection{subsubsection}{3}{-2.45ex}{-3.25ex
plus -1ex minus -.2ex}{1.5ex plus .2ex}{\it }}
\renewcommand{\thesection}{\arabic{section}}
\renewcommand{\thesubsection}{\arabic{section}.\arabic{subsection}}
\renewcommand{\@seccntformat}[1]{\@nameuse{the#1}.~~}

\renewcommand{\theequation}{\thesection.\arabic{equation}}
\makeatletter \@addtoreset{equation}{section}
\def\Ddots{\mathinner{\mkern1mu\raise\p@
\vbox{\kern7\p@\hbox{.}}\mkern2mu
\raise4\p@\hbox{.}\mkern2mu\raise7\p@\hbox{.}\mkern1mu}}
\usepackage[toc,page]{appendix}

\renewcommand{\thethm}{\thesection.\arabic{thm}}

\newcommand{\myxymatrix}[1]{\vcenter{\vbox{\xymatrix{#1}}}}

\renewcommand{\appendices}{
\section*{Appendix}\label{appendices}\setcounter{subsection}{0}
\addcontentsline{toc}{section}{Appendix}
\setcounter{equation}{0}
\makeatletter
\renewcommand{\theequation}{\Alph{subsection}.\arabic{equation}}
\renewcommand{\thesubsection}{\Alph{subsection}}
\renewcommand{\thethm}{\Alph{subsection}.\arabic{thm}}
\@addtoreset{equation}{subsection}
\@addtoreset{thm}{subsection}
\makeatother
}

\def\slasha#1{\setbox0=\hbox{$#1$}#1\hskip-\wd0\hbox to\wd0{\hss\sl/\/\hss}}

\def\periodb#1{\setbox0=\hbox{$#1$}#1\hskip-\wd0\hbox to\wd0{-}}

\newcommand{\unit}{\mathbbm{1}}   			
\newcommand{\id}{\mathrm{id}}   			

\newcommand{\CC}{\mathcal{C}}
\newcommand{\CCC}{\mathscr{C}}
\newcommand{\CCL}{\mathscr{L}}
\newcommand{\CCI}{\mathscr{I}}

\newcommand{\CF}{\mathcal{F}}

\newcommand{\CG}{\mathcal{G}}
\newcommand{\CCG}{\mathscr{G}}
\newcommand{\CH}{\mathcal{H}}
\newcommand{\CCH}{\mathscr{H}}
\newcommand{\CI}{\mathcal{I}}

\newcommand{\CL}{\mathcal{L}}

\newcommand{\CN}{\mathcal{N}}
\newcommand{\CO}{\mathcal{O}}

\newcommand{\CQ}{\mathcal{Q}}

\newcommand{\frg}{\mathfrak{g}}				
\newcommand{\frh}{\mathfrak{h}}				

\newcommand{\FR}{\mathbbm{R}}     			
\newcommand{\FC}{\mathbbm{C}}     			
\newcommand{\NN}{\mathbbm{N}}     			
\newcommand{\RZ}{\mathbbm{Z}}     			

\newcommand{\lambdab}{\bar{\lambda}}

\newcommand{\dd}{\mathrm{d}}     			
\newcommand{\dpar}{\partial}     			
\newcommand{\embd}{{\hookrightarrow}}     		
\newcommand{\di}{\mathrm{i}}     			
\newcommand{\eps}{{\varepsilon}}			
\newcommand{\epsb}{{\bar{\varepsilon}}}			

\newcommand{\sB}{\mathsf{B}}

\newcommand{\eand}{{\qquad\mbox{and}\qquad}}     		
\newcommand{\ewith}{{\qquad\mbox{with}\qquad}}

\newcommand{\der}[1]{\frac{\dpar}{\dpar #1}}   		
\newcommand{\dder}[1]{\frac{\dd}{\dd #1}}   		
\newcommand{\tr}{\,\mathrm{tr}\,}     			

\newcommand{\ab}{\mathfrak{b}}

\newcommand{\au}{\mathfrak{u}}
\newcommand{\asu}{\mathfrak{su}}
\newcommand{\aosp}{\mathfrak{osp}}
\newcommand{\aso}{\mathfrak{so}}
\newcommand{\asp}{\mathfrak{sp}}

\newcommand{\fre}{\mathfrak{e}}
\newcommand{\aspin}{\mathfrak{spin}}
\newcommand{\astring}{\mathfrak{string}}

\newcommand{\sU}{\mathsf{U}}     			

\newcommand{\sG}{\mathsf{G}}

\newcommand{\sL}{\mathsf{L}}

\newcommand{\sSU}{\mathsf{SU}}

\newcommand{\sO}{\mathsf{O}}
\newcommand{\sE}{\mathsf{E}}

\newcommand{\sSO}{\mathsf{SO}}
\newcommand{\sSpin}{\mathsf{Spin}}
\newcommand{\sSp}{\mathsf{Sp}}
\newcommand{\sString}{\mathsf{String}}

\newcommand{\acton}{\vartriangleright}     			
\newcommand{\comment}[1]{}     				
\def\tyng(#1){\hbox{\tiny$\yng(#1)$}}			
\def\tyoung(#1){\hbox{\tiny$\young(#1)$}}			

\newcommand{\langlec}{\prec\hspace{-0.5mm}}
\newcommand{\ranglec}{\hspace{-0.5mm}\succ}

\newcommand{\beq}{\begin{eqnarray}}
\newcommand{\eeq}{\end{eqnarray}}

\definecolor{outrageousorange}{rgb}{1.0, 0.43, 0.29}

\newcommand{\vol}{{\rm vol}}

\begin{document}
\begin{titlepage}
\begin{flushright}
 EMPG--17--23
\end{flushright}
\vskip2.0cm
\begin{center}
{\LARGE \bf Towards an M5-Brane Model I: \\[0.3cm] A 6d Superconformal Field Theory}
\vskip1.5cm
{\Large Christian S\"amann and Lennart Schmidt}
\setcounter{footnote}{0}
\renewcommand{\thefootnote}{\arabic{thefootnote}}
\vskip1cm
{\em Maxwell Institute for Mathematical Sciences\\
Department of Mathematics, Heriot--Watt University\\
Colin Maclaurin Building, Riccarton, Edinburgh EH14 4AS, U.K.}\\[0.5cm]
{Email: {\ttfamily c.saemann@hw.ac.uk~,~ls27@hw.ac.uk}}
\end{center}
\vskip1.0cm
\begin{center}
{\bf Abstract}
\end{center}
\begin{quote}
We present an action for a six-dimensional superconformal field theory containing a non-abelian tensor multiplet. All of the ingredients of this action have been available in the literature. We bring these pieces together by choosing the string Lie 2-algebra as a gauge structure, which we motivated in previous work. The kinematical data contains a connection on a categorified principal bundle, which is the appropriate mathematical description of the parallel transport of self-dual strings. Our action can be written down for each of the simply laced Dynkin diagrams, and each case reduces to a four-dimensional supersymmetric Yang--Mills theory with corresponding gauge Lie algebra. Our action also reduces nicely to an M2-brane model which is a deformation of the ABJM model. While this action is certainly not the desired M5-brane model, we regard it as a key stepping stone towards a potential construction of the (2,0)-theory.
\end{quote}
\end{titlepage}

\tableofcontents

\section{Introduction and results}

Stacks of M5-branes allow for an effective description in terms of a six-dimensional local superconformal quantum field theory, known as the {\em (2,0)-theory}~\cite{Witten:1995zh,Strominger:1995ac,Witten:1995em}. It is a widely held belief that for more than one M5-brane, this theory is non-Lagrangian and exists only at quantum level. In our last paper~\cite{Saemann:2017rjm}, however, we argued that this theory has a classical BPS subsector and that non-abelian self-dual string solitons are captured by a higher gauge theory. In this paper, we present a full classical action of a six-dimensional superconformal field theory which possesses many of the properties expected of an M5-brane analogue of the very successful M2-brane models~\cite{Bagger:2007jr,Gustavsson:2007vu,Aharony:2008ug}. This constitutes an important step towards a potential constructon of a classical action for such an M5-brane analogue. At the very least, our action demonstrates that many of the perceived obstacles to the existence of classical M5-brane models can be overcome.

The (2,0)-theory plays a similarly important role in M-theory as the famous four-dimensional $\CN=4$ super Yang--Mills theory does in string theory. Many aspects of string theory can be understood from the perspective of this gauge theory, its deformations and reductions. An explicit description of the (2,0)-theory would allow us to lift this understanding to M-theory. Moreover, many dualities in string theory have their origin in different compactifications of the (2,0)-theory, which provides a unifying picture. A classical action for this theory would therefore significantly improve our understanding of string and M-theory as a whole. Furthermore, there is considerable scope for applications of the (2,0)-theory within mathematics, see e.g.~\cite{Witten:2009at}.

From more or less strict arguments, one can derive a number of properties that a classical M5-brane model should have, and we review them in more detail in section~\ref{sec:(2,0)-theory}. The most important points on the resulting wish list are the following:
\begin{itemize}
\setlength{\itemsep}{-1mm}
 \item[1)] The action should contain an interacting 2-form gauge potential with self-dual curvature 3-form.
 \item[2)] The action should be based on solid mathematical foundations in order to allow for a formulation on general manifolds.
 \item[3)] The action should have the same field content and moduli space as the $(2,0)$-theory and be at least $\CN=(1,0)$ supersymmetric.
 \item[4)] The gauge structure should arise from Lie algebras of types $A$, $D$ and $E$.
 \item[5)] There should be a restriction of the action to that of a free $\CN=(2,0)$ tensor multiplet.
 \item[6)] The action should have a self-dual string soliton as a BPS state, ideally the one of~\cite{Saemann:2017rjm}.
 \item[7)] There should be an appropriate reduction mechanism to four-dimensional super Yang--Mills theory, yielding gauge Lie algebras of types $A$, $D$ and $E$.
 \item[8)] Ideally, there should be a reduction mechanism to three-dimensional M2-brane models explaining the origin of their discrete Chern--Simons coupling constant.
\end{itemize}

There are a number of objections to the existence of a classical description of the (2,0)-theory, and we shall discuss them in more detail in section~\ref{sec:(2,0)-theory}. A first one has to do with the difficulties in defining parallel transport of extended objects. This problem is addressed by using the mathematically consistent framework of higher gauge theory. Another objection is the absence of continuous coupling constants in the (2,0)-theory. The same, however, is true in the case of M2-branes, and useful M2-brane models have been constructed. A third objection arises from dimensional arguments when considering a dimensional reduction to five dimensions. We circumvent these problems by showing that reductions to four dimensions are possible and yield the expected results.

All of the ingredients in our model have been available in the literature. The gauge sector of our model will be a categorified or {\em higher gauge theory} describing parallel transport of extended objects, since the (2,0)-theory contains 1-dimensional objects and its observables are Wilson surfaces. The mathematical framework of higher principal bundles with connections, which underlies higher gauge theory, has been developed to full extent, see e.g.~\cite{Nikolaus:1207ab} or~\cite{Jurco:2016qwv} and references therein. There is, however, a severe lack of concrete, interesting examples. This appears to be due to the fact that the higher notion of equivalence or isomorphism is very coarse, rendering many constructions trivial. A class of categorified bundles that promises to yield non-trivial results are principal 2-bundles which have as their categorified structure group a 2-group model of the {\em string group}, which is a 3-connected cover of the spin group. Such {\em string structures}~\cite{Killingback:1986rd,Stolz:2004aa,Redden:2006aa,Waldorf:2009uf,Sati:2009ic} appear, in slightly generalized form, in heterotic supergravity and they are also necessary to describe stacks of D-branes when the Kalb--Ramond 2-form $B$ belongs to a topologically non-trivial gerbe. Mathematically, string structures are important e.g.~in the context of elliptic cohomology.

String structures can indeed provide non-trivial and physically relevant examples of higher principal bundles. For instance, they allow for a well-motivated formulation of the non-abelian analogue of the self-dual string soliton~\cite{Saemann:2017rjm}. The relevant equations on $\FR^4$ had been written down as early as in the paper~\cite{Duff:1996cf}, but the full geometric interpretation seems to have remained unclear.

Similarly, a supersymmetric action which secretly allows for generalized string structures as underlying gauge structures has been derived from tensor hierarchies in supergravity~\cite{Samtleben:2011fj,Samtleben:2012mi}. The higher gauge structure hidden in this action was exposed in~\cite{Palmer:2013pka}, see also~\cite{Lavau:2014iva}.\footnote{The gauge structure derived in~\cite{Chu:2011fd} is a particular case of that obtained in~\cite{Samtleben:2011fj}.} The action is conformally invariant and has $\CN=(1,0)$ supersymmetry, which is well-motivated by a comparison with M2-brane models: here only~12 of the expected~16 supercharges are symmetries of the action in general. We shall refer to this action simply as the {\em (1,0)-model} in the following. 

The field content of the (1,0)-model consists of a (1,0)-tensor multiplet as well as a (1,0)-vector multiplet with an involved underlying gauge structure, which can be identified with a refinement of a categorified Lie algebra. To get a model comparable to the ABJM M2-brane model~\cite{Aharony:2008ug}, the action of the (1,0)-model needs to be extended by terms for a (1,0)-hypermultiplet as well as a PST-like term implementing self-duality as an equation of motion. The general coupling to hypermultiplets has been worked out in~\cite{Samtleben:2012fb} and a PST-type extension of the bosonic part of the action was given in~\cite{Bandos:2013jva}. A generalization of the PST extension to the supersymmetric case was announced but has not appeared. It seems to us that this may not be possible in the general gauge algebraic framework used in the (1,0)-model. However, restricting the gauge structure to the string structure in skeletal form, we are able to put all the pieces of the puzzle together and obtain a suitable action. Also some other problems and issues encountered within the (1,0)-model are resolved by this restriction.

Our model, in the special case of higher gauge Lie algebra $\widehat{\astring}(\asu(2))$ and $4\times 4$ hypermultiplets, has the following Lagrangian:
\begin{equation}
\begin{aligned}
\CL &= - 2\partial_\mu\phi_r\partial^\mu\phi_s -8 \bar\chi_r\slasha{\partial}\chi_s-\tfrac{1}{6}\CH_{\mu\nu\kappa}^r\CH_s^{\mu\nu\kappa}+\tfrac16\CH^s_{\mu\nu\kappa}\tr(\bar\lambda\gamma^{\mu\nu\kappa}\lambda)\\
&\phantom{{}={}}-\phi_s\tr\big(F_{\mu\nu}F^{\mu\nu}-2Y_{ij}Y^{ij}+4\bar\lambda\slasha{\nabla}\lambda\big)+4\tr(\bar\lambda F_{\mu\nu})\gamma^{\mu\nu}\chi_s\\
&\phantom{{}={}}-16\tr(Y_{ij}\bar\lambda^i)\chi^j_s+\eps^{\mu\nu\kappa\lambda\rho\sigma}\big(\tfrac{1}{36}C^r_{\mu\nu\kappa}\CH^s_{\lambda\rho\sigma}+\tfrac18B^s_{\mu\nu}\tr(F_{\kappa\lambda}F_{\rho\sigma})\big)\\
&\phantom{{}={}} - \tr(\nabla_\mu q\nabla^\mu q+ 2\bar\psi\slasha{\nabla}\psi - 8\bar\psi [\lambda,q] + 2 q^i[Y_{ij},q^{j}])+\CL_{\rm PST}~.
\end{aligned}
\end{equation}
Here, we have two abelian tensor multiplets $(B_{r,s},\chi^i_{r,s},\phi_{r,s})$, an $\asu(2)$-valued vector multiplet $(A,\lambda^i,Y^{ij})$ and a non-dynamical abelian 3-form field $C_r$ with curvature 2- and 3-forms
\begin{equation}
 F=\dd A+\tfrac12[A,A]\eand \CH=\dd B-(A,\dd A)-\tfrac13(A,[A,A])+C_r~.
\end{equation}
We also have the hypermultiplets $(q^i,\psi)$ taking values in the adjoint representation of $\asu(2)$. The explicit form of the PST term $\CL_{\rm PST}$ is found in~\eqref{eq:Lagrangian_PST}.

This action has all the properties in our wish list for an M5-brane model. It is a mathematically consistent formulation of an interacting 2-form potential and therefore it can be rather readily generalized to arbitrary space-times. It has the same field content as the full $(2,0)$-theory, but only $\CN=(1,0)$ supersymmetry is realized. In section~\ref{ssec:BPS_states}, we briefly comment on the fact that this supersymmetry might be enhanced by self-dual string operators. The gauge structure is rather natural; as discussed in section~\ref{ssec:string_structures}, string structures are the most obvious candidates and they are already implicitly used in many contexts, e.g.~in heterotic supergravity. If we set the $(1,0)$-vector multiplet to zero by choosing the higher gauge Lie algebra $\widehat{\astring}(*)$ and restrict the number of hypermultiplets, we recover the free abelian $\CN=(2,0)$ action. As we shall discuss later, this action has the non-abelian self-dual string soliton of~\cite{Saemann:2017rjm} as a classical BPS configuration. 

In section~\ref{ssec:4d_reduction}, we will show that our action for $\widehat{\astring}(\frg)$ with $\frg$ a Lie algebra of type $A$, $D$ or $E$ straightforwardly restricts to $\CN=2$ super Yang--Mills theory in four dimensions with gauge Lie algebra $\frg$. Even the modulus $\tau$ of the compactifying torus translates into the appropriate couplings, $\tau=\frac{\theta}{2\pi}+\di g_{\rm YM}^{-2}$. Also, a straightforward reduction to an M2-brane model in three dimensions is possible: our action for $\widehat{\astring}(\au(n)\times \au(n))$ turns into a supersymmetric deformation of the ABJM model. In the latter case, the discrete coupling constant is generated as the topological class of the gerbe described by the 2-form potential over the compactifying 3-manifold as explained in section~\ref{ssec:3d_reduction}.

There are, however, a number of crucial discrepancies between our model and the (2,0)-theory which make it clear that this is merely a stepping stone towards an M5-brane model. We plan to address these in future work.

First, the Yang--Mills multiplet in our model contains independent degrees of freedom which are clearly incompatible with $\CN=(2,0)$ supersymmetry. Analogously to the case of M2-brane models, one would expect that these degrees are fixed by a dynamical principle~\cite{Mukhi:2008ux}. 

Second, and related, the moduli space of vacua of our model is not the one expected from a full M5-brane model. In particular, the process of separating individual M5-branes from a stack is not well captured by our model.

Third, the PST mechanism as constructed in~\cite{Bandos:2013jva} relies on a non-vanishing scalar $\phi_s$ in the tensor multiplet. As stated in~\cite{Samtleben:2012fb}, this seems to be related to the tensionless string phase transition~\cite{Seiberg:1996vs,Duff:1996cf}, which, however, is absent under certain conditions. This point requires much further exploration within our model. If this issue (as well as that of the Yang--Mills multiplet) is resolved, there is hope that self-dual string operators might restore full $\CN=(2,0)$ supersymmetry, which should also be studied in detail.

A fourth big issue is a general problem of the (1,0)-model which is not fixed by our choice of gauge structure: There is still a single scalar field with a wrong sign in its kinetic term in the action, and one should find an interpretation for its appearance or a mechanism for its elimination.

Fifth, a physical model based on higher structures should be agnostic with regards to higher isomorphisms. As we argue, the formulation of the model of~\cite{Samtleben:2011fj} is too rigid to allow for this feature. In fact, this can be regarded as something positive, since it provides a clear pointer to possible and necessary extensions of their model.

Sixth, one might be tempted to improve our model such that its dimensional reduction produces the undeformed $\CN=6$ ABJM model. This seems rather inconceivable due to the structure, potentials and dimensions of the matter fields in our model. 

Altogether, we conclude that while our model shares many desired feature with the (2,0)-theory and gives a very useful example of a classical higher gauge theories, it requires further work to qualify as an M5-brane model.

Finally, let us mention that a number of alternative approaches towards finding classical descriptions of the $(2,0)$-theory have been proposed in the literature. A particularly interesting one is to take the M2-brane models as a starting point and to try to reconstruct an M5-brane description. In a first variant~\cite{Ho:2008nn,Ho:2008ve}, a classical M5-brane was derived from an infinite-dimensional 3-Lie algebra based on the diffeomorphisms of the 3-sphere, which are nicely encoded in a Nambu--Poisson 3-Lie algebra\footnote{Note that the 3-Lie algebras underlying the M2-brane models are different from the Lie 3-algebras of higher algebra. The former should be regarded as Lie 2-algebras~\cite{Palmer:2012ya}.}. Such Nambu--Poisson 3-Lie algebras have considerable overlap with categorified Lie algebras, cf.~\cite{Ritter:2015ymv}. In a second variant~\cite{Lambert:2010wm,Lambert:2016xbs}, the steps of the original derivation of M2-brane models were repeated for the M5-brane, leading to 3-Lie algebra valued fields. In both types of models, the underlying mathematical structures are not quite the ones we would expect. This is a severe problem, as global formulations on general manifolds rely on a solid mathematical foundation. Also, it is not clear how these models can fulfill the demands on our above wish list.

Structurally more closely related to our model is the model of~\cite{Ho:2011ni}, see also~\cite{Huang:2012tu,Ho:2014eoa}, which is indeed based on non-abelian gerbes. This model only describes the bosonic gauge part, but it does reduce to five-dimensional gauge theory. It is, however, a theory on $M_5\times S^1$ and contains non-local gauge symmetries.

Finally, there are two approaches that describe the $(2,0)$-theory from a dual perspective. First, there is the paper~\cite{Fiorenza:2012tb}, in which a holographic dual of the $(2,0)$-theory is identified with a higher Chern--Simons theory and which uses essentially the same language and structures we are employing in this paper. Second, there is an approach based on the twistor description of the $(2,0)$-theory's equations of motion~\cite{Saemann:2011nb,Saemann:2012uq,Saemann:2013pca,Jurco:2014mva,Jurco:2016qwv}. As already suspected in~\cite{Lambert:2010wm}, these equations might exist classically, even if the corresponding Lagrangian did not. If this is the case, then there may be a twistor description for these, just as there is one for the $\CN=3$ super Yang--Mills equations. In the latter case, holomorphic bundles over a $5|6$-dimensional supertwistor space yield solutions to the $\CN=3$ super Yang--Mills equations. Since the twistor space for self-dual 3-form in six dimensions is known, this reduces the search for a (2,0)-theory effectively to the problem of defining an appropriate higher gauge structure: The latter leads, rather unambiguously, to higher bundles over this twistor space, which yield solutions to constraint equations for a higher superconnection on $\FR^{1,5|16}$. These constraint equations are manifestly $\CN=(2,0)$ supersymmetric and conformally invariant and they are equivalent to certain equations of motion on the field content of the $(2,0)$-tensor multiplet. An appropriate choice of higher gauge structure for rendering these equations interacting in an interesting way has, however, not yet been identified, but the twisted string structures of our model should be the right ingredient.

We hope to be able to report on results tying at least some of these alternative perspectives to our model in future work.

This paper is structured as follows. In section~\ref{sec:(2,0)-theory}, we give a short review of the (2,0)-theory, highlighting desirable properties of a classical M5-brane model as well as arguments against its existence. Section~\ref{sec:tools} provides a concise summary of definitions, literature references and context for the mathematical tools that we employ in section~\ref{sec:action} to construct our model. Its dimensional reductions to super Yang--Mills theory, M2-brane models and $L_\infty$-algebra models are discussed in section~\ref{sec:dim_reductions}.

\section{The (2,0)-theory in six dimensions}\label{sec:(2,0)-theory}

Among the supersymmetric field theories, superconformal ones are of particular interest because of their simplicity and the role they play in many areas of string theory, cf.~\cite{Moore:2012aa}. While a conformal field theory on the pseudo-Riemannian manifold $\FR^{p,q}$ is invariant under an action of the conformal algebra $\aso(p+1,q+1)$, a superconformal theory has to be invariant under an action of a super extension of this Lie algebra. Such extensions only exist for $p+q\leq 6$~\cite{Nahm:1977tg}, see also~\cite{Shnider:1988wh}. Explicit examples of conformal (and superconformal) field theories have been known for a long time, and it was suspected that four was the maximal dimension for non-trivial unitary conformal field theories. However, string theory and M-theory strongly suggest that there should be a non-trivial six-dimensional superconformal field theory and it is generally believed today that the highest dimensional superconformal algebra also underlies an interesting quantum field theory.

In particular, type IIB string theory on $\FR^{1,5}\times$K3 has a moduli space of vacua which is a homogeneous space with orbifold singularities classified by the simply laced Dynkin diagrams of types $A$, $D$ and $E$. At these points, the volume of a 2-cycle in the K3 vanishes, and near such a singularity, a D3-brane wrapping the corresponding 2-cycle turns into a non-critical string in six dimensions~\cite{Witten:1995zh,Seiberg:1996vs}. The mass of this string is related to the volume of the 2-cycle, which is proportional to the distance to the singularity in the moduli space. This string is self-dual in the sense that the 3-form $H=\dd B$ it produces is self-dual in $\FR^{1,5}$. Also, these strings are massless at the singularity and they decouple from all other string modes. This suggests that there is a self-contained and consistent quantum field theory describing these {\em self-dual strings}. 

There is also an M-theory interpretation of self-dual strings in terms of boundaries of M2-branes ending on parallel M5-branes, where the former mediate the interactions of the latter~\cite{Strominger:1995ac,Dasgupta:1995zm,Witten:1995em}. As the M5-branes approach each other, the self-dual strings become massless and we recover the above picture. Concretely, $N$ flat M5-branes correspond to the (2,0)-theory at an $A_{N-1}$-singularity, while the $D$-series arises when several M5-branes come together at an $\FR^5/\RZ_2$-singularity. The $E$-series seems to be less clearly understood.

The link between M-theory and type IIB interpretations can be made more explicit~\cite{Dasgupta:1995zm,Witten:1995em}. A further connection between M-theory and the six-dimensional (2,0)-theory was provided in the initial paper on the AdS/CFT-correspondence~\cite{Maldacena:1997re}, where a duality was conjectured between M-theory on AdS$_7\times S^4$ and the $(2,0)$-theory; see also~\cite{Fiorenza:2012tb}.

The theory of massless self-dual strings has been argued to be a local quantum field theory~\cite{Seiberg:1996qx}. In particular, it is an $\CN=(2,0)$ superconformal field theory in six-dimensions and the superconformal Lie algebra of symmetries is $\aosp(6,2|4)$ containing the conformal algebra $\aso(6,2)$ and the R-symmetry Lie algebra $\aso(5)\cong \asp(2)$. From~\cite{Nahm:1977tg} we know that the relevant representation of this algebra is the $(2,0)$-tensor multiplet, whose bosonic part consists of a self-dual 3-form field strength and five scalars. The latter can be regarded as the Goldstone scalars for the breaking of the symmetry group $\sSO(1,10)$ of $\FR^{1,10}$ to $\sSO(1,5)\times \sSO(5)$ due to the presence of a flat M5-brane. Correspondingly, they describe the position of the M5-brane in the five directions orthogonal to the M5-brane's worldvolume. The degrees of freedom should scale as $\CO(N^3)$ with the number $N$ of M5-branes.

The observables of the (2,0)-theory are Wilson loops~\cite{Ganor:1996nf}, see also e.g.~\cite{Corrado:1999pi,Chen:2007ir}. This suggests that a classical description underlying the (2,0)-theory should be a theory of parallel transport of the self-dual strings: Just as the effective description of D-branes in terms of ordinary gauge theory captures the parallel transport of the endpoints of strings on them, an effective description of M5-branes should capture the parallel transport of the boundaries of the M2-branes on them. The mathematical description of such a higher-dimensional parallel transport is called {\em higher gauge theory}, cf.~\cite{Baez:2010ya,Saemann:2016sis}, and it is geometrically based on categorified notions of Lie groups, Lie algebras and principal bundles. In the abelian case, categorified principal bundles are also known as gerbes. Rather less well-known seems to be the fact that there is also a complete mathematical theory of non-abelian gerbes, see~\cite{Nikolaus:1207ab} and references therein. 

Ordinary gerbes are recognized to play an important part in string theory since the Kalb--Ramond 2-form field $B$ is, globally, part of the connective structure of a gerbe~\cite{Gawedzki:1987ak,Freed:1999vc}. There is, however, some reluctance to appreciate non-abelian versions of gerbes even though these have been rediscovered and used in many contexts. For example, the non-abelian gerbes relevant to our discussion are also the global geometric object underlying heterotic string theory as well as the description of stacks of D-branes and they become non-trivial when the $B$-field corresponds to a topologically non-trivial 3-form flux~$H$.

The various no-go theorems in the literature which concern non-abelian higher gauge theory can be circumvented as we shall explain in section~\ref{ssec:higher_gauge_theory}. A valid point of critique of this theory, however, is the small number of non-trivial and physically relevant examples and applications. This may be attributed to the rapid development of this very young field. Also, the ease with which generalizations are made using category theory makes development of the general theory more appealing than the rather cumbersome process of working out specific examples. On closer inspection, and with some guidance from string theory, it is however not too hard to find examples. In particular, a consistent and well-motivated description of non-abelian analogues of self-dual strings using the framework of higher gauge theory was given in~\cite{Saemann:2017rjm}.

The prejudice against non-abelian higher gauge theory is certainly one reason for the general belief that the (2,0)-theory has no description in terms of a classical Lagrangian, cf.~\cite{Witten:2007ct}. Another, more relevant argument goes as follows: The (2,0)-theory is a conformal field theory, so it does not have any dimensionful parameters. Moreover, the singularities in the moduli space at which the theory of self-dual strings becomes conformal are isolated. Therefore, there are no continuous dimensionless parameters either. This clearly suggests that there is no classical limit and therefore no Lagrangian description. Note that the fact that there is no continuous parameter that can act as a coupling constant and thus as a continuous deformation of the abelian theory was also derived by computing the appropriate BRST cohomology group~\cite{Bekaert:9909094,Bekaert:2000qx}.

We can simply address this concern by pointing to the successful construction of M2-brane models~\cite{Bagger:2007jr,Gustavsson:2007vu,Aharony:2008ug} since very similar arguments are valid in their case. Note that in  M2-brane models, the Chern--Simons coupling was given by a discrete geometric parameter \mbox{$k\in\NN$} arising from an orbifold $\FC^4/\RZ_k$, circumventing the lack of continuous dimensionless parameters. We do expect an analogous argument to apply to the case of M5-branes.

This parallel suggests also another point about classical descriptions, previously observed in~\cite{Samtleben:2011fj}. Just as there are no general $\CN=8$ supersymmetric M2-brane models\footnote{once certain reasonable conditions are imposed} but only ones with $\CN=6$ supersymmetry, we should merely expect $\CN=(1,0)$ supersymmetric M5-brane models and none with full $\CN=(2,0)$ supersymmetry.

A final argument that makes the existence of a Lagrangian description of the six-dimensional (2,0)-theory implausible stems from dimensional reduction~\cite{Witten:2007ct,Witten:2009at}: We know that the six-dimensional (2,0)-theory should reduce to maximally supersymmetric Yang--Mills theory in five dimensions after compactification on a circle of radius $R$, which leads to a volume form $2\pi R\,\dd^5 x$ in the action. On the other hand, conformal invariance in six dimensions as well as dimensional analysis of the Yang--Mills term in the Lagrangian requires a volume form $\frac{1}{R}\,\dd^5 x$. We think that this is a valid point; however, we show that a direct and classical reduction to four dimensions can be performed in section~\ref{ssec:4d_reduction}. Combined with some dimensional oxidation, this can lead to the five-dimensional super Yang--Mills theory, albeit in an indirect fashion.

Having addressed all common arguments against a Lagrangian description, let us continue with the discussion of some of the features we should expect. 

An M5-brane model should be able to capture the process of separating individual or stacks of M5-brane from another stack. In particular, it should be able to describe the Coulomb branch given by a separation of all individual M5-branes from each other. This is perhaps the most difficult property to model since it is unclear, already at a mathematical level, how an analogue of the branching $\sU(n)\rightarrow \sU(1)^{\times n}$ relevant for D-branes should work in the case of categorified Lie groups.

Furthermore, the $(2,0)$-theory on a manifold $M_6=M_4\times E$, with $E$ an elliptic curve, dimensionally reduces to $\CN=4$ gauge theory on $M_4$ with complex coupling constant $\tau$ determined by $E$. The type of the Dynkin diagram classifying the K3-singularity determines the gauge algebra of the four-dimensional theory, an important feature we wish to reproduce. Indeed, the $(2,0)$-theory can be used to gain insights into four-dimensional gauge theories and their BPS states, cf.~\cite{Witten:1995zh,Klemm:1996bj}.

Similarly, there may be a reduction of the $(2,0)$-theory to the above mentioned M2-brane models. In such a reduction, one would have to explain in particular the origin of the discrete Chern--Simons coupling.

As in the case of supersymmetric Yang--Mills theories, interesting classical configurations of the (2,0)-theory would be given by BPS states, which are also known as {\em self-dual string solitons}, see also~\cite{Dijkgraaf:1996hk} and~\cite{Howe:1997ue}. We should expect that these BPS states also feature in our classical Lagrangian description. 

Altogether, we arrive at the wish list of features given in the introduction.

\section{Mathematical tools}\label{sec:tools}

In this section, we concisely summarize the mathematical tools relevant for discussing a higher gauge theory in order to be reasonably self contained and to provide relevant references. Readers familiar with this material may directly jump to section~\ref{sec:action}, others might want to consult also the reviews~\cite{Baez:2010ya,Saemann:2016sis} and in particular the papers~\cite{Sati:2008eg,Sati:2008kz,Sati:2009ic,Fiorenza:2010mh,Nikolaus:1207ab}.

\subsection{Higher gauge theory}\label{ssec:higher_gauge_theory}

Ordinary gauge theory describes the parallel transport of point particles and higher gauge theory is the natural analogue for extended objects~\cite{Baez:2010ya,Baez:2002jn}. This is of central interest in string and M-theory given that their fundamental objects are higher dimensional. Similarly, it will be crucial in our description of self-dual strings.

Parallel transport of ordinary point particles transforming under a gauge group $\sG$ assigns to each path a group element in $\sG$. Mathematically, this gives rise to the holonomy functor from the path groupoid of the base manifold to $\sB \sG=(*\leftleftarrows \sG)$, the one-object groupoid of the gauge group $\sG$. For the parallel transport of strings we should then, analogously, assign some group data for each surface swept out by the string. Such an assignment has to satisfy various consistency conditions. In particular, the different ways of composing the parallel transports in the diagram
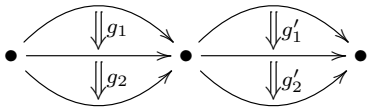
\begin{equation*}
 \myxymatrix{
\bullet\ar@/_4ex/[rr]_{}="g1"
  \ar[rr]_(0.65){}
  \ar@{}[rr]|{}="g2"
  \ar@/^4ex/[rr]^{}="g3"
  \ar@{=>}^{g_2} "g2"+<0ex,-0.5ex>;"g1"+<0ex,+0.5ex>
  \ar@{=>}^{g_1} "g3"+<0ex,-0.5ex>;"g2"+<0ex,0.5ex>&& 
\bullet\ar@/_4ex/[rr]_{}="h1"
  \ar[rr]_(0.65){}
  \ar@{}[rr]|{}="h2"
  \ar@/^4ex/[rr]^{}="h3"
  \ar@{=>}^{g'_2} "h2"+<0ex,-0.5ex>;"h1"+<0ex,0.5ex>
  \ar@{=>}^{g'_1} "h3"+<0ex,-0.5ex>;"h2"+<0ex,0.5ex>
&& \bullet
}
\end{equation*}
should agree. In equations, this means that
\begin{equation}
(g_1 g_2)(g'_1g'_2)=(g_1g'_1)(g_2g'_2)~,
\end{equation}
which forces $\sG$ to be abelian, due to an argument by Eckmann and Hilton~\cite{Hilton:1962:227-255}. This argument was rediscovered, in infinitesimal form, e.g.\ in~\cite{Teitelboim:1985ya}. However, there is a well-known generalization of this equation in 2-categories with their two ways of composing morphisms, $\otimes$ and $\circ$, known as the \textit{interchange law}:
\begin{equation}
(g_1\circ g_2)\otimes(g'_1\circ g'_2) = (g_1\otimes g'_1)\circ(g_2\otimes g'_2)~.
\end{equation}
This is also natural from the point of view of parallel transport along surfaces: we have 1-morphisms, denoted by $\rightarrow$, together with 2-morphisms, denoted by $\Rightarrow$, which are morphisms between morphisms. Also, parallel transport becomes a 2-functor. 

The gauge group is analogously replaced by a {\em Lie 2-group}, which is a monoidal category with invertible morphism and objects that are weakly invertible with respect to the monoidal product. A detailed discussion of a rather general notion of Lie 2-groups, their properties and a classification result is found in~\cite{Baez:0307200}.\footnote{Note, however, that the string 2-group model of~\cite{Schommer-Pries:0911.2483}, which becomes relevant when globalizing our model, requires a more general definition of Lie 2-groups than that given in this reference.} 

There is a Lie functor differentiating Lie 2-groups to Lie 2-algebras, where the latter are given in the form of the $L_\infty$-algebras familiar from BRST/BV quantization as well as string field theory~\cite{Severa:2006aa}, see also~\cite{Jurco:2014mva}. Since we shall be exclusively interested in the local description of the (2,0)-theory over contractible patches in this paper, infinitesimal symmetries are sufficient for our purposes and we can ignore Lie 2-groups and turn to the somewhat simpler Lie 2-algebras.

Before giving the precise definition of these in section~\ref{subsec:Linfty}, let us briefly outline the global geometric picture underlying higher gauge theory, which is encoded in higher principal bundles with connection. To arrive at definitions of categorified principal bundles we replace all notions in the various definitions of principal bundles by more or less straightforwardly categorified analogues: A \textit{smooth 2-space} is a category internal to $\mathsf{Diff}$, the category of smooth manifolds with smooth morphisms. Note that any manifold $M$ naturally gives rise to the trivial smooth 2-space $M\leftleftarrows M$ in which all morphism are identities. A \textit{smooth $\CCG$-space} is a smooth 2-space on which we have a smooth action of the Lie 2-group $\CCG$. A \textit{principal $\CCG$-bundle} over a manifold $M$ is then a locally trivial $\CCG$-space over the 2-space $M\leftleftarrows M$~\cite{Wockel:2008aa}. 

An equivalent viewpoint is given by looking at the transition functions of principal $\sG$-bundles for $\sG$ some Lie group. These are described via functors from the \textit{\v Cech groupoid} $\CCC(Y)=(Y\leftleftarrows Y^{\left[2\right]} )$ of a surjective submersion $\sigma: Y \twoheadrightarrow M$ to the groupoid $\sB\sG$. This readily extends to the categorified case: here the transition functions arise as 2-functors from the \v Cech groupoid, trivially viewed as a 2-category, to the delooping $\sB \CCG$ of the 2-group $\CCG$, which in turn give rise to a principal $\CCG$-bundle~\cite{Wockel:2008aa}. For $\CCG = \sB \sU(1)$, we recover abelian gerbes or principal $\sB\sU(1)$-bundles. Topologically, these are characterized by an element of $H^3(M,\RZ)$, called the \textit{Dixmier-Douady class}. This is the analogue of the first Chern class of line bundles, which is an element in $H^2(M,\RZ)$.

The parallel transport of particles and strings on a principal $\CCG$-bundle is then encoded in a connection with associated curvature. The connection can either be described by global forms on the total 2-space of the principal 2-bundle, see e.g.~\cite{Waldorf:1608.00401}, or by local forms over patches of the base 2-space, which are glued together by the cocycles describing the principal 2-bundle, see e.g.~\cite{Jurco:2014mva}.

For the local description, there exists an elegant and useful framework of higher gauge theory, where these connection and curvature forms are described in terms of morphism of differential graded algebras. This formalism allows to deduce the relevant local data defining a connection, the appropriate corresponding notions of curvature forms, the gauge transformations as well as the Bianchi identities.\footnote{Local expressions that globalize to topological invariants (i.e.~higher analogues of Chern classes) are also readily derived.} Thus, the full local kinematical data of higher gauge theory results from a morphism of differential graded algebras. For a concise review, see e.g.~\cite{Saemann:2017rjm}.

Let us stress that up to technicalities in global descriptions, these constructions readily generalize to the parallel transport of membranes and even higher dimensional objects. In fact, our model is based on a principal 4-bundle, an extension necessary to describe a generalized self-duality condition in six dimensions.

\subsection{Categorified Lie algebras: \texorpdfstring{$L_\infty$}{L-infinity}-algebras}\label{subsec:Linfty}

The infinitesimal symmetries underlying higher gauge theory are captured by categorified Lie algebras. For all our purposes, these can be regarded as equivalent to $L_\infty$-algebras~\cite{Lada:1992wc,Lada:1994mn}, which also underlie closed string field theory~\cite{Zwiebach:1992ie}. 

Let us begin with {\em Lie 2-algebras}. Just as a Lie algebra has an underlying vector space, a Lie 2-algebra $\CCL$ has an underlying categorified vector space $\CCL=(\CCL_0\leftleftarrows\CCL_1)$. The most appropriate definition of a 2-vector space is still under dispute, but sufficient for us is a category in which both objects and morphisms are vector spaces and all maps are linear. In addition, we have a Lie bracket functor $[\cdot,\cdot]:\CCL\times \CCL\to \CCL$ and a natural isomorphism, the Jacobiator, that relaxes the Jacobi identity and satisfies some higher coherence axiom~\cite{Baez:2003aa}. As shown in that paper, Lie 2-algebras are categorically equivalent to 2-term $L_\infty$-algebras, and we now switch to this more convenient and familiar language.

{\em Strong homotopy Lie algebras}, or {\em $L_\infty$-algebras} for short, are given by a $\RZ$-graded vector space $\sL = \sum_{k\in \RZ} \sL_k$ together with a set of totally antisymmetric, multilinear maps \mbox{$\mu_i : \wedge^i \sL \to \sL,~i\in \NN,$} of degree $i-2$, which satisfy the \textit{higher} or \textit{homotopy} Jacobi relations
\begin{equation}\label{eq:hom_rel}
\sum\limits_{i+j=n} \sum\limits_\sigma (-1)^{ij} \chi(\sigma; \ell_1,\dots,\ell_n) \mu_{j+1}(\mu_i(\ell_{\sigma(1)},\dots,\ell_{\sigma(i)}),\ell_{\sigma(i+1},\dots,\ell_{\sigma(n)}) = 0
\end{equation}
for all $n\in\NN$ and $\ell_1,\dots,\ell_n\in \sL$, where the second sum runs over all $(i,j)$-\textit{unshuffles}, that is, permutations whose image consists of ordered sets of length $i$ and $j$: $\sigma(1)<\dots<\sigma(i)$ and $\sigma(i+1)<\dots<\sigma(i+j)$. Additionally, $\chi(\sigma; \ell_1,\dots,\ell_n)$ denotes the \textit{graded Koszul sign} defined by the graded antisymmetrized products
\begin{equation}
\ell_1\wedge\dots \wedge \ell_n = \chi(\sigma;\ell_1,\dots,\ell_n)\ell_{\sigma(1)}\wedge\dots\wedge \ell_{\sigma(n)}~,
\end{equation}
where any transposition not involving only odd degree elements acquires a minus sign. An \textit{$n$-term $L_\infty$-algebra} is an $L_\infty$-algebra that is concentrated (i.e.~non-trivial only) in degrees $0,\dots,n-1$. 

In this paper we are mostly interested in Lie 2- and 3-algebras. The lowest homotopy Jacobi relations read as follows:
\begin{equation}
\begin{aligned}
0&=\mu_1\left(\mu_1\left(\ell_1\right)\right)~,\\
0&= (-1)^{\left| \ell_1\right|  \left| \ell_2\right| } \mu _2\left(\mu _1\left(\ell_2\right),\ell_1\right)+\mu _1\left(\mu _2\left(\ell_1,\ell_2\right)\right)-\mu _2\left(\mu _1\left(\ell_1\right),\ell_2\right)~,\\
0 &=(-1)^{\left| \ell_2\right|  \left| \ell_3\right| +1} \mu _2\left(\mu _2\left(\ell_1,\ell_3\right),\ell_2\right)+(-1)^{\left| \ell_1\right|  \left(\left| \ell_2\right| +\left| \ell_3\right|
   \right)} \mu _2\left(\mu _2\left(\ell_2,\ell_3\right),\ell_1\right)\\
&\phantom{{}={}}+(-1)^{\left| \ell_1\right|  \left| \ell_2\right| +1} \mu _3\left(\mu _1\left(\ell_2\right),\ell_1,\ell_3\right)+(-1)^{\left(\left| \ell_1\right|
   +\left| \ell_2\right| \right) \left| \ell_3\right| } \mu _3\left(\mu _1\left(\ell_3\right),\ell_1,\ell_2\right)\\
&\phantom{{}={}}+\mu _1\left(\mu _3\left(\ell_1,\ell_2,\ell_3\right)\right)+\mu _2\left(\mu
   _2\left(\ell_1,\ell_2\right),\ell_3\right)+\mu _3\left(\mu _1\left(\ell_1\right),\ell_2,\ell_3\right)~,
\end{aligned}
\end{equation}
where $\ell_i \in \sL$. These relations show that $\mu_1$ is a graded differential compatible with $\mu_2$, and $\mu_2$ is a generalization of a Lie bracket with the violation of the Jacobi identity controlled by $\mu_3$.

An elegant and particularly useful way of describing $L_\infty$-algebras is a dual description in terms of differential graded algebras arising from functions on $Q$-manifolds, which also feature prominently in BRST and BV quantization. A \textit{$Q$-manifold} is a $\RZ$-graded manifold $M$ together with a \textit{homological vector field} $Q$, i.e.~a vector field of degree~1 satisfying \mbox{$Q^2=0$}. Now, given an $n$-term $L_\infty$-algebra $\sL$ we can grade-shift the individual homogeneously graded vector spaces by~1 yielding the graded manifold
\begin{equation}
\sL[1] = (*\leftarrow \sL_0[1] \leftarrow\dots\leftarrow \sL_{n-1}[1])~.
\end{equation}
The degree of the maps $\mu_i$ is accordingly shifted from $i-2$ to $-1$, and this allows one to define the codifferential
\begin{equation}
D = \sum\limits_i \mu_i~,
\end{equation}
which has degree $-1$ and acts on the graded symmetric tensor algebra ${\rm Sym}^\bullet(\sL[1])$. Dualizing to the algebra of functions on $\sL[1]$ yields a corresponding differential $Q$ of degree $1$. Thus, $\sL[1]$ becomes a $Q$-manifold, and we arrive at a differential graded algebra ${\rm CE}(\sL)=(\CC^\infty(\sL[1]),Q)$, known as the \textit{Chevalley-Eilenberg algebra} of $\sL$. The condition $Q^2=0$ corresponds precisely to the homotopy Jacobi relations~\eqref{eq:hom_rel}. 

In order to write down action principles for gauge theories, we use mostly metric matrix Lie algebras with the metric given by the trace. More generally, there is a compatibility relation between metric and Lie bracket,
\begin{equation}
 (a,[b,c])=(c,[a,b])~.
\end{equation}
In the case of $L_\infty$-algebras, we need an analogous concept. Such a \textit{cyclic structure} on an $L_\infty$-algebra $\sL$ over $\FR$ is a graded symmetric, non-degenerate bilinear form
\begin{equation}
\left\langle -,-\right\rangle\,:\, \sL\odot \sL \to \FR
\end{equation}
satisfying the following compatibility condition for all $\ell_i \in \sL$:
\begin{equation}
\left\langle \ell_1,\mu_i(\ell_2,\dots,\ell_{i+1})\right\rangle = (-1)^{i+|\ell_{i+1}|(|\ell_1|+\dots+|\ell_i|)}\left\langle \ell_{i+1},\mu_i(\ell_1,\dots,\ell_i)\right\rangle~.
\end{equation}
That is, we can cyclically permute the $\ell_i$ while respecting the usual Koszul convention for permuting graded elements. 

\subsection{The string Lie 2-algebra}\label{ssec:string_Lie_2}

Non-trivial examples of Lie 2-algebras which are interesting and relevant to physical applications are rare. The most prominent ones are perhaps the abelian Lie 2-algebra $\ab\au(1)=(*\leftarrow \au(1))$ underlying infinitesimal gauge transformation on abelian gerbes as well as the resulting semidirect product with infinitesimal diffeomorphisms encoded in the famous exact Courant algebroid $TM\oplus T^*M$ for some manifold $M$, see~\cite{Deser:2016qkw} and references therein. This lack of useful examples is perhaps one of the key reasons for the skepticism that higher gauge theory meets in applications to string theory.

However, a small set of reasonable examples can also be an advantage (as long as the set is not empty): it suggest some form of uniqueness. Indeed, from certain perspectives it seems that the Lie 2-algebra associated to the exact Courant algebroid, its Lie 2-subalgebras and combinations thereof are essentially all the examples that are needed: $\ab\au(1)$ is included, as is the string Lie 2-algebra $\astring(n)$, cf.~\cite{Baez:2009:aa}. The latter will be the one underlying our action. In the following, we briefly outline its origin from a Lie 2-group model of the string group $\sString(n)$. For more details, see~\cite{Baez:2003aa,Baez:2005sn,Saemann:2017rjm}.

Recall that the spin group $\sSpin(n)$ fits into a sequence of Lie groups, known as the {\em Whitehead tower} of $\sO(n)$, which is linked by group homomorphisms inducing isomorphisms on all homotopy groups except for the lowest non-trivial one. The string group $\sString(n)$ is simply an element in the sequence
\begin{equation}
 \dots \rightarrow \sString(n)\rightarrow \sSpin(n)\rightarrow \sSpin(n)\rightarrow \sSO(n)\rightarrow \sO(n)~,
\end{equation}
rendering it a 3-connected cover of $\sSpin(n)$. This sequence defines $\sString(n)$ only up to certain equivalences, which leads to a variety of models of the string group. Particularly useful ones are models as Lie 2-groups, and a first one was given in~\cite{Baez:2005sn}. This model is a strict Lie 2-group $\sString_{\hat\Omega}(n)$ consisting of the based path space as well as a Kac--Moody central extension of the based loop group of $\sSpin(n)$. A second model which is finite dimensional but relies on Segal--Mitchison cohomology was presented in~\cite{Schommer-Pries:0911.2483}. It can be Lie differentiated to a Lie 2-algebra of the following form~\cite{Demessie:2016ieh}:
\begin{equation}
 \astring_{\rm sk}(n)\ =\ \big(~\aspin(n)~\xleftarrow{~0~}~\FR~\big)
\end{equation}
with non-trivial products
\begin{equation}\label{eq:string_Lie_2_brackets}
\begin{aligned}
 \mu_2&:\aspin(n)\wedge \aspin(n)\rightarrow \aspin(n)~,~~~&\mu_2(x_1,x_2)&=[x_1,x_2]~,\\
 \mu_3&:\aspin(n)\wedge \aspin(n)\wedge \aspin(n)\rightarrow \FR~,~~~&\mu_3(x_1,x_2,x_3)&=(x_1,[x_2,x_3])~,
\end{aligned}
\end{equation}
where $(-,-)$ is the Cartan--Killing form on $\aspin(n)$. Clearly, such a Lie 2-algebra exists for any metric Lie algebra $\frg$, in particular, for those of types $A$, $D$, and $E$, and we write $\astring_{\rm sk}(\frg)$ for the corresponding Lie 2-algebra with non-trivial Lie 2-algebra products~\eqref{eq:string_Lie_2_brackets}. 

Note that the string Lie 2-algebra itself does not carry a cyclic structure as defined in the previous section; this is only possible for Lie 2-algebras $\sL_0\leftarrow \sL_1$ with $\sL_0\cong \sL_1$. We do, however, need such a structure for defining an action, and we therefore have to work with an extension of the string Lie 2-algebra which we shall develop in section~\ref{ssec:gauge_structure}.

\section{A six-dimensional superconformal field theory}\label{sec:action}

In this section, we use the formalism summarized in the previous section to combine the constructions of~\cite{Samtleben:2011fj,Samtleben:2012fb,Bandos:2013jva} into an interesting $\CN=(1,0)$ superconformal action, which satisfies all the criteria collected in the wish list of the introduction. We start with an outline of the ingredients before we discuss the action, its supersymmetries and the equations of motion.

\subsection{Context}\label{ssec:intro_to_action_construction}

Our starting point is the $\CN=(1,0)$ superconformal model presented in~\cite{Samtleben:2011fj}, which we simply call the {\em (1,0)-model}. The (1,0)-model was obtained by writing down an ansatz for suitable supersymmetry transformations for a non-abelian tensor multiplet and deriving algebraic and dynamic conditions for their closure. This is the same method that led to the BLG M2-brane model~\cite{Bagger:2007jr,Gustavsson:2007vu}. 

The gauge structure of the (1,0)-model and its gauge field contents were derived from the non-abelian tensor hierarchy, see e.g.~\cite{deWit:2008ta} and references in~\cite{Samtleben:2011fj}. Maximal supergravities can be constructed by compactifying 10- and 11-dimensional supergravities on a torus. Each maximal supergravity exhibits a duality group $\sG$, and the theory can be deformed by rendering the one-form gauge potentials non-abelian with the gauge group being a subgroup of the duality group $\sG$. The precise structure is encoded in a representation of the subgroup of $\sG$ and the embedding tensor $\Theta$. The algebraic structure yields covariant derivatives and field strengths which, however, may not transform covariantly. In such cases, one is led to introduce a compensating non-abelian 2-form gauge potential, whose 3-form curvature may also not transform covariantly. This, in turn, forces the introduction of a compensating non-abelian 3-form potential and so on.

As explained in~\cite{Palmer:2013pka}, the algebraic structures underlying the non-abelian tensor hierarchy are categorified Lie algebras and the iterative construction of higher form potentials leads essentially to the same formulas for curvatures, gauge transformations and Bianchi identities as found in higher gauge theory, see also~\cite{Fiorenza:2012tb} and~\cite{Lavau:2014iva}. The (1,0)-model provides therefore a useful starting point for our constructions. 

To achieve our goal, we have to address a few issues with the action of the (1,0)-model. First, we have to complete the field content to contain that of the full $(2,0)$-theory, even though we are just looking for an $\CN=(1,0)$ superconformal action. This is analogous to the ABJM model, which is only $\CN=6$ supersymmetric, but has the same field content as the full $\CN=8$ BLG model. For this, we can rely on the results of~\cite{Samtleben:2012fb}. Second, we would like to incorporate the PST formalism~\cite{Pasti:1995ii,Pasti:1995tn} to include self-duality of the 3-form curvature as an equation of motion of the action. For the bosonic part of the $(1,0)$-model, this has been done in~\cite{Bandos:2013jva}. An extension to the supersymmetric case was announced, but this has not appeared yet in the literature. It is actually not clear to us, that such a construction is possible in the general gauge structure considered in~\cite{Samtleben:2011fj}. In fact, the details of our PST mechanism seem to differ in some points from those of~\cite{Bandos:2013jva}.

Third, we have to specify a useful, more explicit form of the gauge structure of the (1,0)-model. It is here where we draw on results from higher gauge theory, in particular our discussion in~\cite{Saemann:2017rjm}: For each metric Lie algebra, we construct a string Lie 2-algebra model that can be extended to a Lie 3-algebra and yields a suitable gauge structure for the (1,0)-model as explained in section~\ref{ssec:gauge_structure}. This will also solve most of the issues with the (1,0)-model and its PST extension: The cubic interactions will vanish and the PST and hypermultiplet extensions are rather straightforward.

\subsection{String structures}\label{ssec:string_structures}

As argued in detail in the literature, the most obvious candidate for the appropriate Lie 2-algebra describing multiple M5-branes is the string Lie 2-algebra. Let us collect and summarize a few of these arguments before detailing the algebraic structure. Recall that the gauge potential on stacks of D-branes in a topologically non-trivial $B$-field background is described by a connection on a twisted principal bundle. In~\cite{Aschieri:2004yz}, the higher analogue for M-theory was identified with twisted gerbes, and the relevant gauge group was given by the central extension of the based loop space of $E_8$. Also, appropriate degrees of freedom have been derived from a boundary ABJM model in~\cite{Chu:2009ms}, which form a $\sU(2N)\times \sU(2N)$ Kac--Moody current algebra. The Kac--Moody central extension of the based loop group of $E_8$, is indeed the important ingredient in the string 2-group model for $\sE_8$ of~\cite{Baez:2005sn}. A much more explicit and extended version of this argument is found in~\cite{Fiorenza:2012tb}.

Also, as argued in~\cite{Saemann:2017rjm} within the context of non-abelian extensions of the self-dual string soliton, the string Lie 2-algebra of $\aspin(3)\cong \asu(2)$ is the direct higher analogue of $\asu(2)$ in various ways. Most importantly, just as the manifold underlying the Lie group $\sSpin(3)\cong \sSU(2)\cong S^3$ is the total space of the fundamental circle bundle (the Hopf fibration) over $S^2$, the categorified space underlying the Lie 2-group $\sString(3)$ is the total space of the fundamental abelian gerbe over $S^3$.

Further arguments stem e.g.~from expectations such as higher analogues of the fuzzy funnel which opens when D1-branes end on D3-branes. The worldvolume of the D1-branes polarizes into fuzzy 2-spheres, whose corresponding Hilbert spaces carry representations of the double cover $\sSpin(3)$ of the isometry group $\sSO(3)$. Similarly, we expect that the worldvolume of M2-branes ending on M5-branes polarizes into fuzzy 3-spheres. The latter should be quantized in a categorified way as discussed in~\cite{Bunk:2016rta}. As argued in that paper, the string group acts naturally on the categorified Hilbert spaces expected in a categorified quantization of $S^3$.

A last argument for using the string Lie 2-algebra is simply the lack of suitable alternatives. The appropriate notion of equivalence between Lie 2-algebras is that of quasi-isomorphisms, and we expect that correctly formulated physical applications do not make a distinction between quasi-isomorphic Lie 2-algebras. This, however, renders most naively constructed examples of Lie 2-algebras either equivalent to ordinary Lie algebras or essentially abelian.

A {\em string structure} is basically a categorified principal bundle whose structure Lie 2-group is a 2-group model of the string group~\cite{Killingback:1986rd,Stolz:2004aa,Redden:2006aa,Waldorf:2009uf,Sati:2009ic}. We shall always consider string structures carrying a categorified connection. This geometrical data underlies heterotic supergravity as well as the gauge theory description of stacks of multiple D-branes in the background of a $B$-field belonging to a topologically non-trivial gerbe.

The global picture of such a bundle in terms of cocycle data is readily derived for the strict 2-group model of~\cite{Baez:2005sn}; for the model of~\cite{Schommer-Pries:0911.2483}, the global description was given in~\cite{Demessie:2016ieh}. In the following, we shall focus on the local description and work over the contractible space $\FR^{1,5}$.

The relevant data here was identified long ago in the context of heterotic supergravity~\cite{Bergshoeff:1981um,Chapline:1982ww}. In $d$ dimensions, we start from the gauge algebra $\frg\coloneqq \aspin(d)\times \frh$, where $\frh$ should be thought of as one of the gauge Lie algebras $\fre_8\times \fre_8$ or $\aso(32)$. Both Lie algebras $\aspin(d)$ and $\frh$ are endowed with the Cartan--Killing form.\footnote{If we want to homogenize the formulas in the following in order to work with $\frg$, we have to invert the sign of the Cartan--Killing form on $\frh$.} We introduce the gauge potential 1-forms $\omega$ and $A$, taking values in $\aspin(d)$ and $\frh$, respectively. Additionally we have a potential 2-form $B$ taking values in $\au(1)$. The appropriate curvature 2- and 3-forms of these gauge potentials read as
\begin{subequations}\label{eq:twisted_string_structures}
\begin{equation}\label{eq:gg_fs}
 F_\omega=\dd \omega+\tfrac12 [\omega,\omega]~,~~~F_A=\dd A+\tfrac12[A,A]~,~~~H=\dd B+{\rm cs}(\omega)-{\rm cs}(A)~,
\end{equation}
where ${\rm cs}(A)=(A,\dd A)+\tfrac13(A,[A,A])$ is the usual Chern--Simons form. 

Note that the 3-form gauge potential is a 3-form taking values in $\au(1)$. This might be one reason why higher gauge theory with the string Lie 2-algebra as gauge Lie 2-algebra has been underappreciated: This situation is too familiar from heterotic supergravity and does not seem like a truly non-abelian version of higher gauge theory at first glance. Nevertheless, a slight extension of string structures will prove to be rich enough for all our purposes.

Infinitesimal gauge transformations of string structures are parameterized by functions $\alpha_0$ and $\alpha_1$ with values in $\aspin(d)$ and $\frh$, respectively, as well as a $\au(1)$-valued 1-form $\Lambda$. They act according to 
\begin{equation}
\begin{aligned}
 \delta \omega &= \dd \alpha_0 + [\omega,\alpha_0]~,\\
 \delta A &= \dd\alpha_1 + [A,\alpha_1]~,\\
 \delta B &= \dd \Lambda +(\alpha_0,\dd \omega)-(\alpha_1,\dd A)~.
\end{aligned}
\end{equation}
These gauge transformations leave $H$ invariant. The higher Bianchi identities are readily found to read as 
\begin{equation}\label{eq:gg_Bianchi}
 \dd F_\omega+[\omega,F_\omega]=0~,~~~\dd F_A+ [A,F_A]=0~,~~~\dd H=(F_\omega,F_\omega)-(F_A,F_A)~.
\end{equation}
\end{subequations}
Note that the last identity is the Green--Schwarz anomaly cancellation condition~\cite{Green:1984sg}. 

The above formulas for gauge transformations, curvatures and Bianchi identities differ from those canonically derived in higher gauge theory, see e.g.\ the discussion in~\cite{Saemann:2017rjm}. A comprehensive mathematical interpretation of these was given in~\cite{Sati:2009ic}, see also~\cite{Sati:2008eg}, and we refer to those papers, as well as the review in~\cite{Saemann:2017rjm}, for a more detailed explanation of string structures and their twist. 

This form of kinematical data of higher gauge theory is also recovered, in a slightly generalized form, from the discussion of tensor hierarchies, cf.\ section~\ref{ssec:intro_to_action_construction}. Therefore, it provides the higher gauge structure underlying the (1,0)-model of~\cite{Samtleben:2011fj}, which will be the foundation of our action.

As a final remark, recall that global string structures are usually defined with the additional constraint that the first Pontrjagin class of $F_\omega+F_A$ vanishes.

\subsection{The higher gauge algebra}\label{ssec:gauge_structure}

Following the arguments of the previous section, we start from the string-like  Lie 2-algebra $\astring_{\rm sk}(\frg)=(~\frg~\leftarrow~\FR~)$ with non-trivial brackets~\eqref{eq:string_Lie_2_brackets}. Since the total space of $\astring_{\rm sk}(\frg)$ is not a symplectic $Q$-manifold, it does not carry a cyclic structure, the appropriate form of an inner product for $L_\infty$-algebras. We therefore have to minimally extend this Lie 2-algebra to a cotangent space. One possibility would be the Lie 2-algebra $ \frg \oplus \FR^*\leftarrow\frg^*\oplus \FR$. 

A better solution is motivated by the fact that the gauge structure underlying the (1,0)-model is indeed a Lie 3-algebra refined by additional structures as explained in~\cite{Palmer:2013pka} and later in~\cite{Lavau:2014iva}. Moreover, string structures are most conveniently defined starting from a Lie 3-algebra quasi-isomorphic to $\aspin(n)$~\cite{Sati:2008kz,Sati:2009ic}. This suggests to use a Lie 3-algebra first considered in~\cite{Saemann:2017rjm}: Given a metric Lie algebra $\frg$, this Lie 3-algebra is 
\begin{equation}
 \sL=\big(\frg\oplus \FR^*\leftarrow \FR[1]\oplus \FR^*[1] \leftarrow \frg^*[2]\oplus \FR[2]\big)~,
\end{equation}
where for a vector space $W$, $W[i]$ again denotes the elements in $W$ shifted in degree by $i$. The natural symplectic form $\omega$ on the $Q$-manifold given by the graded vector space
\begin{equation}
\sL[1]~~=~~\frg[1]\oplus \FR^*[1]~~\oplus~~\FR[2]\oplus \FR^*[2]~~\oplus~~\frg^*[3]\oplus \FR[3]
\end{equation}
reads as 
\begin{equation}
 \omega=\dd \xi^\alpha\wedge \dd \zeta_\alpha+\dd q\wedge \dd p+\dd s\wedge \dd r
\end{equation}
in terms of coordinates $(\xi^\alpha, q)$ of degree~1, $(r, s)$ of degree~2 and $(\zeta_\alpha, p)$ of degree~2. The homological vector field $Q$ is an extension of that on the string Lie 2-algebra $\astring_{\rm sk}(\frg)$ and its Hamiltonian $\CQ$ with $Q=\{\CQ,-\}$ reads as
\begin{equation}
 \CQ=-\tfrac12 f^\alpha_{\beta\gamma} \xi^\beta\xi^\gamma\zeta_\alpha-\tfrac{1}{3!}f_{\alpha\beta\gamma}\xi^\alpha\xi^\beta\xi^\gamma s+s p~,
\end{equation}
where $f_{\alpha\beta\gamma}$ are the structure constants on $\frg$ with an index lowered by the metric. 

Translating this back to bracket notation, we have a cyclic inner product on $\sL$,
\begin{multline}
 \langle \xi_1+q_1+r_1+s_1+p_1+\zeta_1,\xi_2+q_2+r_2+s_2+p_2+\zeta_2\rangle=\\\zeta_1(\xi_2)+\zeta_2(\xi_1)+p_1(q_2)+p_2(q_1)+r_1(s_2)+r_2(s_1)
\end{multline}
with $\xi_{1,2}+q_{1,2}\in \frg\oplus \FR^*$, $r_{1,2}+s_{1,2}\in \FR[1]\oplus \FR^*[1]$ and $\zeta_{1,2}+p_{1,2}\in \frg^*[2]\oplus \FR[2]$ and $p_1(q_2)=p_1q_2$ etc. The non-trivial higher products of the Lie 3-algebra $\sL$ are 
\begin{equation}\label{eq:higher_products_standard_Lie_3}
\begin{aligned}
 &\mu_1:\FR^*[1]\rightarrow \FR^*:~\mu_1(s)\coloneqq s~,\\
 &\mu_1:\FR[2]\rightarrow \FR[1]:~\mu_1(p)\coloneqq p~,\\
 &\mu_2:\frg\wedge \frg\rightarrow \frg:~ \mu_2(\xi_1,\xi_2)\coloneqq [\xi_1,\xi_2]~,\\
 &\mu_2:\frg\wedge \frg^*[2]\rightarrow \frg^*[2]:~ \mu_2(x,y)\coloneqq y([-,x])~,\\
 &\mu_3:\frg\wedge \frg\wedge \frg\rightarrow \FR[1]:~\mu_3(\xi_1,\xi_2,\xi_3)\coloneqq (\xi_1,[\xi_2,\xi_3])~,\\
 &\mu_3:\frg\wedge \frg\wedge \FR^*[1]\rightarrow \frg^*[2]:~\mu_3(\xi_1,\xi_2,s)\coloneqq \langle(-,[\xi_1,\xi_2]),s\rangle~.
\end{aligned}
\end{equation}
Furthermore, we have the following maps of degree~1, refining the Lie 3-algebra structure:
\begin{equation}\label{eq:higher_products_refined_Lie_3}
\begin{aligned}
 &\nu_2:\frg\otimes \frg \rightarrow \FR[1]: \nu_2(\xi_1,\xi_2):=(\xi_1,\xi_2)~\mbox{from the metric on $\frg$}~,\\
 &\nu_2:\frg\otimes \FR^*[1]\rightarrow \frg^*[2]: \nu_2(\xi,s)\coloneqq 2\langle\nu_2(-,\xi),s\rangle~.
\end{aligned}
\end{equation}
For all other arguments, the maps $\mu_i$ and $\nu_2$ vanish.

This Lie 3-algebra is sufficient to write down an action. In order to encode the full duality relations for differential forms in six dimensions, however, we have to add a non-propagating four-form. We therefore have to extend this Lie 3-algebra trivially to a Lie 4-algebra as follows:\footnote{One can always extend a Lie $n$-algebra to a Lie $n+1$-algebra by adding the kernel of $\mu_1$. This is usually problematic, as it renders the Lie $n+1$-algebra quasi-isomorphic to a Lie $n-1$-algebra. Here, however, the resulting Lie 4-algebra is refined by additional structure which avoids this problem.}
\begin{equation}
\xymatrixcolsep{4pc}
\xymatrixrowsep{1pc}
\myxymatrix{
 \FR^*\ar@{}[d]|{\oplus}\ar@{<-}[r]^{\mu_1=\id} & \FR^*[1]\ar@{}[d]|{\oplus} & \frg^*[2]\ar@{}[d]|{\oplus}\ar@{<-^{)}}[r]^{\mu_1=\id} & ~\frg^*[3]\\
 \frg& \FR[1]\ar@{<-}[r]^{\mu_1=\id} & \FR[2] & }
\end{equation}
This is the diagram to have in mind when we will discuss the field content and the action below. The only additional maps we define are 
\begin{equation}\label{eq:higher_products_extension_Lie_4}
\begin{aligned}
 &\mu_2:\frg\wedge \frg^*[3]\rightarrow \frg^*[3]: \mu_2(x,z)=z([-,x])~,\\
 &\nu_2:\frg\otimes \frg^*[2]\rightarrow \frg^*[3]: \nu_2(\xi,\zeta)\coloneqq \zeta([\xi,-])~.
\end{aligned}
\end{equation}
We denote the resulting refined Lie 4-algebra simply by $\widehat{\astring}(\frg)$.

The maps above indeed encode a gauge structure for a (1,0)-model as discussed in detail in~\cite{Saemann:2017rjm}. The explicit dictionary to the structure constants used in~\cite{Samtleben:2011fj} is the following:
\begin{center}
\begin{tabular}{@{}lll@{}}
\toprule
     & Notation~\cite{Samtleben:2011fj,Bandos:2013jva} & Translated to $\widehat\astring(\frg)$\\
\midrule
   Indices & $T^r$ & $T^\alpha+T_q\in \frg\oplus \FR^*$\\
   ($T$: general obj.)& $T^I$ & $T_r+T_s \in \FR[1]\oplus \FR^*[1]$\\
   & $T_r$ & $T_\alpha+T_p\in \frg^*[2]\oplus \FR[2]$\\
   & $T_\alpha$ & $T_\alpha\in \frg^*[3]$\\
\midrule
   Structure const.\ & $h^r_I$ & $\mu_1=\id:\FR^*[1]\rightarrow \FR^*$\\
   & $g^{Ir}$ & $\mu_1=\id:\FR[2]\rightarrow \FR[1]$\\
   & $k_r^\alpha$ & $\mu_1=\id:\frg^*[3]\rightarrow \frg^*[2]$\\
   & $f_{st}{}^r$ & $\mu_2:\frg\wedge \frg\rightarrow \frg:~ \mu_2(\xi_1,\xi_2)\coloneqq [\xi_1,\xi_2]$\\
   & $d_{rs}^I$ & $-\nu_2:\frg\otimes \frg \rightarrow \FR[1]: -(\xi_1,\xi_2)$\\
   & $b_{Irs}$ & $-\nu_2:\frg\otimes \FR^*[1]\rightarrow \frg^*[2]: -\nu_2(\xi,s)\coloneqq -2\langle(-,\xi),s\rangle$\\
   & $c^t_{\alpha s}$ & $-\nu_2:\frg\otimes \frg^*[2]\rightarrow \frg^*[3]: -\nu_2(\xi,\zeta)\coloneqq -\zeta([\xi,-])$\\
\bottomrule
 \end{tabular}
\end{center}
\vspace*{0.5cm}
\phantom{a}

\subsection{Kinematical data: Gauge sector}\label{ssec:kin_data_gauge_sector}

The relevant field content of our action contains the categorified connection on a principal 4-bundle with structure Lie 4-algebra $\widehat{\astring}(\frg)$, whose field strengths are defined similarly to those of string structures. 

Given a metric Lie algebra $\frg$, the categorified connection is given by the fields
\begin{equation}
\begin{aligned}
 A&\in\Omega^1(\FR^{1,5})\otimes (\frg\oplus \FR^*)~,~~~&B&\in \Omega^2(\FR^{1,5})\otimes (\FR[1]\oplus \FR^*[1])~,\\
 C&\in \Omega^3(\FR^{1,5})\otimes (\frg^*[2]\oplus\FR[2])~,~~~&D&\in \Omega^4(\FR^{1,5})\otimes \frg^*[3]~,
\end{aligned}
\end{equation}
where $D$ will be a non-propagating 4-form potential. Note that the difference between form degree and Lie 4-algebra degree is always~1, cf.~e.g.~the discussion in~\cite{Jurco:2014mva}. The corresponding curvatures read as
\begin{equation}\label{eq:higher_curvs}
\begin{aligned}
 \CF&=\dd A+\tfrac12 \mu_2(A,A)+\mu_1(B)&~~&\in~~\Omega^2(\FR^{1,5})\otimes (\frg\oplus\FR^*)~,\\
 \CH&=\dd B-\nu_2(A,\dd A)-\tfrac13\nu_2(A,\mu_2(A,A))+\mu_1(C)&~~&\in~~\Omega^3(\FR^{1,5})\otimes (\FR[1]\oplus\FR^*[1])~,\\
 \CG&=\dd C+\mu_2(A,C)+\nu_2(\CF,B)+\mu_1(D)&~~&\in~~\Omega^4(\FR^{1,5})\otimes (\frg^*[2]\oplus\FR[2])~,\\
 \CI&=\dd D+\nu_2(\CF,C)+\dots&~~&\in~~\Omega^5(\FR^{1,5})\otimes \frg^*[3]~,
\end{aligned}
\end{equation}
where the definitions of all the relevant maps are found in~\eqref{eq:higher_products_standard_Lie_3},~\eqref{eq:higher_products_refined_Lie_3} and~\eqref{eq:higher_products_extension_Lie_4} and the remaining terms in $\CI$ are of no importance to our discussion. If familiar with such structures, one clearly recognizes an extension of the string structure to a connective structure on a categorified principal bundle with a structure Lie 4-algebra.\footnote{The full construction of this extension would lead us to far from the main topic of this paper. We plan to develop this in an upcoming publication.} The total degree of the curvature forms is always~2. 

The Bianchi identities are readily computed to take the following form:
\begin{equation}\label{eq:bianchi_higher_curvs}
\begin{aligned}
 \nabla \CF-\mu_1(\CH)&=0~,~~~&\dd \CH+\nu_2(\CF,\CF)-\mu_1(\CG)&=0~,\\
 \nabla \CG-\nu_2(\CF,\CH)-\mu_1(\CI)&=0~,~~~&\nabla \CI+\nu_2(\CF,\CG)+\dots&=0~.
\end{aligned}
\end{equation}
Here, $\nabla$ is the covariant derivative
\begin{equation}
 \nabla \phi\coloneqq \dd \phi+\mu_2(A,\phi)
\end{equation}
for any field $\phi$ taking values in $\widehat{\astring}(\frg)$. The combination of $\nabla\pm\mu_1$ as a differential operator of total degree~1 is very natural in higher gauge theory, cf.~e.g.~the last section of~\cite{Demessie:2014ewa}. The additional terms involving $\nu_2$ are due to using string structure-like curvatures instead of the canonical ones. This also motivates the introduction of the following generalized notions of variation, cf.~\cite{Samtleben:2011fj}:
\begin{equation}
 \Delta B\coloneqq\delta B-\nu_2(\delta A,A)~,~~~\Delta C\coloneqq\delta C+\nu_2(\delta A,B)~,~~~\Delta D\coloneqq\delta D-\nu_2(\delta A,C)~,
\end{equation}
which allows us to write
\begin{equation}\label{eq:gen_var_higher_curvs}
\begin{aligned}
\delta \CF&=\nabla \delta A+\mu_1(\Delta B)\\
\delta \CH&=\dd (\Delta B)-2\nu_2(\CF,\delta A)+\mu_1(\Delta C)~,\\
\delta \CG&= \nabla (\Delta C) + \nu_2(\delta A, \CH) +\nu_2(\CF,\Delta B) + \mu_1(\Delta D)~,\\
\delta \CI&= \dd\delta D+\nu_2(\CF,\delta C)+ \nu_2(\nabla\delta A,C)+\dots~.
\end{aligned}
\end{equation}

Infinitesimal gauge transformations are parameterized by 
\begin{equation}
\begin{aligned}
 \alpha&\in \Omega^0(\FR^{1,5})\otimes (\frg\oplus \FR^*)~,&~~~\Lambda&\in\Omega^1(\FR^{1,5})\otimes (\FR[1]\oplus \FR^*[1])~,\\\Sigma&\in \Omega^2(\FR^{1,5})\otimes (\frg^*[2]\oplus \FR[2])~,&~~~\Xi&\in \Omega^3(\FR^{1,5})\otimes \frg^*[3]~,
\end{aligned}
\end{equation}
and they modify gauge potentials and their curvatures as follows:
\begin{equation}
 \begin{aligned}
  \delta A&=\dd \alpha + \mu_2(A,\alpha)-\mu_1(\Lambda) ~,&~~~\delta \CF&=\mu_2(\CF,\alpha)~,\\
  \delta B&=\dd \Lambda +\nu_2(\CF,\alpha)-\tfrac12\mu_3(A,A,\alpha)-\mu_1(\Sigma) ~,&~~~\delta \CH&=0~,\\
  \delta C&=\dd \Sigma +\mu_2(C,\alpha)+\mu_2(A,\Sigma)+\nu_2(\CF,\Lambda)-\mu_1(\Xi) ~,&~~~\delta \CG&=\mu_2(\CG,\alpha)~,\\
  \delta D&= \dd \Xi -\nu_2(\CF,\Sigma)+\dots ~,&~~~\delta \CI&=\mu_2(\CI,\alpha)~.
 \end{aligned}
\end{equation}

\subsection{Kinematical data: Supersymmetry partners}\label{ssec:kin_data_susy_partners}

Let us now complete the above extension of a string structure by adding the remaining fields of the full $\CN=(2,0)$ tensor supermultiplet and introduce $\CN=(1,0)$ superpartners for the 1-form gauge potential. We use the same fields and spinor conventions as in~\cite{Samtleben:2011fj,Samtleben:2012fb}, see also~\cite{Bergshoeff:1985mz}.

The R-symmetry group for $\CN=(1,0)$ supersymmetry is $\sSp(1)$, and therefore all fields arrange in representations of this group. We use $i,j$ as indices for representations of $\sSp(1)$. R-symmetry indices are raised and lowered using the Levi--Civita symbol $\eps^{ij}$ and its inverse $\eps_{ij}$: $\lambda_i=-\eps_{ij} \lambda^j$ and $\lambda^i=\eps^{ij}\lambda_j$. Also, we abbreviate NW-SE contractions of indices: $\bar \lambda\psi=\bar\lambda^i\psi_i$.

First, we have the vector supermultiplet containing the one-form gauge potential $A$, a doublet of symplectic Majorana--Weyl spinors $\lambda^i$, satisfying $\gamma_7\lambda^i=\lambda^i$, as well as a triplet of auxiliary scalar fields $Y^{ij}=Y^{(ij)}$, all taking values in $\frg\oplus \FR^*$. Supersymmetry transformations are parameterized by a doublet of chiral spinor $\eps^i$ with $\gamma_7\eps^i=\eps^i$ and read as
\begin{equation}\label{eq:susy_trafo_vector}
 \begin{aligned}
  \delta A&=-\bar \eps \gamma_{(1)} \lambda~,\\
  \delta \lambda^i&=\tfrac{1}{8}\gamma^{\mu\nu}\CF_{\mu\nu}\eps^i~,\\
  \delta Y^{ij}&=-\bar \eps^{(i}\gamma^\mu \nabla_\mu \lambda^{j)}+2\mu_1(\bar\eps^{(i}\chi^{j)})~,
 \end{aligned}
\end{equation}
where we used the notation $\gamma_{(p)}=\dd x^{\mu_1}\wedge \dots \wedge \dd x^{\mu_p}\gamma_{\mu_1}\dots\gamma_{\mu_p}$. We also suppressed all evident R-symmetry index contractions.

Infinitesimal gauge transformations, parameterized by $(\alpha,\Lambda,\Sigma,\Xi)$, act on the additional fields in the vector multiplet according to
\begin{equation}
 \delta \lambda=\mu_2(\lambda,\alpha)\eand \delta Y^{ij}=\mu_2(Y^{ij},\alpha)~.
\end{equation}

Second, we have the tensor supermultiplet containing the two-form gauge potential $B$, a doublet of Majorana--Weyl spinors $\chi^i$ satisfying $\gamma_7\chi^i=-\chi^i$ and a single scalar field $\phi$, all taking values in $\FR[1]\oplus \FR^*[1]$. The supersymmetry transformations read as
\begin{equation}\label{eq:susy_trafo_tensor}
 \begin{aligned}
  \delta \phi&=\bar \eps \chi~,\\
  \delta \chi^i&=\tfrac{1}{48}\gamma^{\mu\nu\rho} \CH_{\mu\nu\rho}\eps^i+\tfrac{1}{4}\slasha{\dd}\phi\eps^i+\tfrac12 \nu_2(\gamma^\mu\lambda^i,\bar \eps \gamma_\mu \lambda)~,\\
  \Delta B&=-\bar \eps \gamma_{(2)}\chi~.
 \end{aligned}
\end{equation}
Note that gauge transformations act trivially on the fields $\chi$ and $\phi$.

We also have the following supersymmetry transformation for the 3-form potential $C$:
\begin{equation}
 \Delta C=\nu_2(\bar \eps \gamma_{(3)}\lambda,\phi)~.
\end{equation}

Next, we come to the matter fields which form hypermultiplets. A detailed review of the general situation is found in~\cite{Samtleben:2012fb}. In the following, we only repeat what is necessary for our construction which has a flat target space.

We start by embedding our gauge Lie algebra $\frg$ into $\asp(n)$. Recall that the group $\sSp(n)\cong \mathsf{USp}(2n)$ is given by $2n\times 2n$-dimensional unitary complex matrices $m$ such $m^T\Omega m=\Omega$ and we choose
\begin{equation}
 \Omega=\begin{pmatrix}
         0 & \unit_n \\ -\unit_n & 0
        \end{pmatrix}
\end{equation}
for the symplectic form. Therefore the Lie algebra $\asp(n)=\mathfrak{usp}(2n)$ consists of complex block matrices
\begin{equation}\label{eq:form_usp2n}
 u=\begin{pmatrix}
    A & B \\ -B^* & - A^T
   \end{pmatrix}
\ewith A^\dagger=-A\eand B^T=B
\end{equation}
and as a vector space, it has dimension $n^2+n(n+1)=n(2n+1)$. By putting $B=0$ in~\eqref{eq:form_usp2n}, we obtain an embedding $\au(n)\embd \asp(n)$. Demanding that $A$ is a real matrix leads to an embedding $\aso(n)\embd \asp(n)$ and considering subgroups of $\au(n)$ leads to the $E$-series. Altogether, we can indeed embed any of the Lie algebras of types $ADE$ into $\asp(n)$, and we denote the generators of the original Lie algebra $\frg$ in this matrix embedding by $u_\alpha{}^a{}_b$, where again $\alpha=1,\dots,\dim (\frg)$ and $a,b=1,\dots, 2n$.

We are particularly interested in the cases $\frg=\asu(N)$ and $\frg=\au(N)\times \au(N)$ and we shall embed both cases into $\au(N^2)\subset \asp(N^2)$ to obtain adjoint and bifundamental representations\footnote{Recall that this can be done via vectorization and the Kronecker product: given matrices $A,B,C,D$, we have
\begin{equation}
 ABC=D~~\Leftrightarrow (C^T\otimes A){\rm vec}(B)={\rm vec}(D)~,
\end{equation}
where ${\rm vec}(B)$ is the vector consisting of the columns of $A$ stacked on top of each other. In particular, if $\lambda_\alpha$ is a generator of $\asu(N)$ in the fundamental representation, then $\unit\otimes \lambda_\alpha-\lambda_\alpha^T\otimes \unit$ is the corresponding generator in the vectorization of the adjoint.}. Correspondingly, we have $\FR^{2\times 2N^2}$ scalar fields encoded in $2\times 2n$-dimensional matrices $q^{ia}$, $i=1,2$, $a=1,\dots,2n$, where the index $i$ labels a vector of the R-symmetry group $\sSp(1)$. The superpartners of the scalar fields are $2n$ antichiral, symplectic Majorana spinors $\psi^a$, satisfying $\gamma_7\psi=-\psi$.

This choice of the number of hypermultiplets arises from doubling the degrees of freedom in the $\asu(N)$-valued $(1,0)$-vector multiplet and the $\au(1)$-valued $(1,0)$-tensor multiplet that we have in our theory. More justification arises from the various dimensional reductions discussed in section~\ref{sec:dim_reductions}.

We now define the obvious action of the gauge Lie algebra $\frg$ on $\FR^{2n}$ by
\begin{equation}
 \acton: \frg\times \FR^{2n}\rightarrow \FR^{2n}~,~~~((\xi^\alpha\tau_\alpha)\acton x)^a=\xi^\alpha u_\alpha{}^a{}_b x^b~,
\end{equation}
where $\tau_\alpha$ are the generators of $\frg$. We shall also use the bilinear pairing
\begin{equation}
 \langlec -,-\ranglec: \FR^{2n}\times \FR^{2n}\rightarrow \FR~,~~~\langlec x,y\ranglec:=\Omega_{ab}x^a y^b
\end{equation}
for all $x,y\in \FR^{2n}$. Infinitesimal gauge transformations, parameterized again by $(\alpha,\Lambda,\Sigma,\Xi)$, act then on the fields in the hypermultiplets as
\begin{equation}
 \delta q^i= \alpha\acton q^i\eand\delta \psi=\alpha \acton \psi~,
\end{equation}
and the covariant derivatives are given by
\begin{equation}
  \nabla_\mu q^i:=\dpar_\mu q^i+A \acton q^i\eand \nabla_\mu \psi:=\dpar_\mu \psi+A \acton \psi~.
\end{equation}
The supersymmetry transformations for the fields in the hypermultiplets read as
\begin{equation}
 \begin{aligned}
  \delta q^{ia}&=\epsb^i\psi^a\eand \delta \psi^a&=\tfrac12 \slasha{\nabla}q^{ia}\eps_i~.
 \end{aligned}
\end{equation}

Finally, we introduce duality invariant supersymmetric extensions of the curvature forms~\eqref{eq:higher_curvs}, which will become convenient later:
\begin{equation}\label{eq:super_higher_curvatures}
\begin{aligned}
 \CCH&:=*\CH-\CH-\nu_2(\bar \lambda,\gamma_{(3)}\lambda)~,\\
 \CCG&:=\CG-\nu_2(*\CF,\phi)+2\nu_2(\bar \lambda,*\gamma_{(2)}\chi)~,\\
 \CCI&:=\CI+\mu_2(\nu_2(\bar\lambda,\phi),\gamma^\mu\lambda)\vol_\mu+2\langlec-\acton q,*\nabla q\ranglec-2\langlec\bar\psi,-\acton\gamma^\mu\psi\ranglec\vol_\mu~,
\end{aligned}
\end{equation}
where $\vol_\mu=\iota_{\der{x^\mu}}\vol$ is the contraction of the volume form on $\FR^{1,5}$ by $\der{x^\mu}$ and terms like $\langlec-\acton q,*\nabla q\ranglec$ denote elements of $\frg^*[3]$. We choose a convention such that $\gamma_{(3)}$ is anti-self-dual.

\subsection{Dynamics: Action}

Our action is composed of ingredients collected from~\cite{Samtleben:2011fj,Samtleben:2012fb,Bandos:2013jva} and consists of four parts: 
\begin{equation}\label{eq:full_action}
 S=\int_{\FR^{1,6}}~\CL_{\rm tensor}~+~\CL_{\rm top}~+~\CL_{\rm hyper}~+~\CL_{\rm PST}~.
\end{equation}
We shall now explain these terms in detail, using the notation, maps and fields defined in sections~\ref{ssec:gauge_structure},~\ref{ssec:kin_data_gauge_sector} and~\ref{ssec:kin_data_susy_partners}.

The first part, $\CL_{\rm tensor}$, consists of the terms coupling the $(1,0)$-tensor multiplet to the $(1,0)$-vector multiplet and reads as
\begin{equation}
\begin{aligned}
\CL_{\rm tensor} &= - \langle\dd \phi,*\dd \phi\rangle -4\vol \langle\bar\chi,\slasha{\dd}\chi\rangle-\tfrac12\langle\CH,*\CH\rangle+\langle\CH,\nu_2(\bar\lambda,*\gamma_{(3)}\lambda)\rangle\\
&\phantom{{}={}} -2\big\langle\phi\,,\,\nu_2(\CF,*\CF)-2\vol~\nu_2(Y_{ij},Y^{ij})+4\vol~\nu_2(\bar\lambda,\slasha{\nabla} \lambda)\big\rangle \\
&\phantom{{}={}}+8\big\langle \nu_2(\bar\lambda,\CF),*\gamma_{(2)}\chi\big\rangle -16\vol\big\langle \nu_2(Y_{ij},\bar\lambda^i),\chi^j\big\rangle~.
\end{aligned}
\end{equation}
Most of this action is expected, given our gauge structure. The term $\langle \phi,\nu_2(\CF,*\CF)\rangle$, e.g., was essentially used already in~\cite{Seiberg:1996qx}, and $\CL_{\rm tensor}$ contains its supersymmetric completion.

The second part, $\CL_{\rm top}$, is a complementing topological term, 
\begin{equation}
\CL_{\rm top} = \langle\mu_1(C),\CH\rangle +\big\langle B,\nu_2(\CF,\CF)\big\rangle~.
\end{equation}
This topological term is due to the presence of the additional, Chern--Simons-like terms in the curvatures~\eqref{eq:higher_curvs}. It can also be seen as arising from the boundary contribution of a manifestly gauge invariant 7-form given by
\begin{equation}
\dd \CL_{\rm top} = \langle\mu_1(\CG),\CH\rangle+\langle\CH,\nu_2(\CF,\CF)\rangle~.
\end{equation}

The third part, $\CL_{\rm hyper}$, contains the kinetic and coupling terms for the $(1,0)$-hyper multiplet:
\begin{equation}
\CL_{\rm hyper} = -\langlec \nabla q,*\nabla q\ranglec + 2\vol\langlec\bar\psi,\slasha{\nabla}\psi\ranglec + 8\vol\langlec\bar\psi, \lambda_i\acton q^i\ranglec + 2\vol\,\langlec q^i,Y_{ij}\acton q^j\ranglec~.
\end{equation}
This part of the Lagrangian can be multiplied with any factor without breaking the supersymmetry. Here, we normalize such that the kinetic terms of $q$ and $\phi$ have the same coefficient, as would be the case if the $\sSpin(5)=\sSp(2)$ R-symmetry of the (2,0)-theory was realized.

Finally, the PST mechanism which lets the self-duality of $\CH$ appear as an equation of motion is implemented by adding the last part, $\CL_{\rm PST}$. In order to be explicit, let us introduce the pairing
\begin{equation}
\begin{aligned}
 \Phi:\frg^*[2]\oplus \FR[2]\rightarrow \frg\oplus \FR^*,~~~\Phi(\zeta+p)&:=\frac{1}{\phi_s}(\zeta,-)~,
\end{aligned}
\end{equation}
where $(-,-):\frg^*\times \frg^*\rightarrow \FR^*$ is the inverse of the metric $(-,-)$ on $\frg$ and $\phi_s:=\phi|_{\FR^*[1]}$. Clearly, $\Phi$ is only defined if $\phi_s\neq 0$, and this is the first time we encounter the tensionless string phase transition. The PST term of the action then reads as 
\begin{equation}\label{eq:Lagrangian_PST}
\CL_{\rm PST} = \tfrac12\big\langle\iota_{V}\CCH,\CCH\big\rangle\wedge v+\langle \Phi(\iota_{V} * \CCG), *~\iota_{V}*\CCG\rangle~,
\end{equation}
where the duality invariant supersymmetrically extended higher curvatures $\CCH$ and $\CCG$ were defined in~\eqref{eq:super_higher_curvatures}, $v$ is a nowhere vanishing exact auxiliary one-form and $V$ its corresponding dual vector field:
\begin{equation}
v = v_\mu \dd x^\mu = \dd a,~~~\iota_V v=1~,~~~\iota_V *v=0
\end{equation}
for some auxiliary scalar field $a$. These additional terms allow for a manifestly Lorentz-invariant Lagrangian that includes the expected duality equations in its equations of motion without having to impose these by hand, see~\cite{Pasti:1995ii,Pasti:1995tn} for original references and, furthermore,~\cite{Pasti:1996vs,Pasti:1997gx,Bandos:9701149,Aganagic:9701166} for follow-ups. 

\subsection{Dynamics: Equations of motion}\label{ssec:dynamics_pst}

Since we extend the purely bosonic computations of~\cite{Bandos:2013jva} to the supersymmetric case and because our use of the Lie 4-algebra $\widehat{\astring}(\frg)$ introduces simplifications, let us present the PST mechanism in a little more detail. Before starting we list a few identities that prove very useful in all subsequent calculations. First, let us note that
\begin{equation}
 \langle\nu_2(\bar\lambda,\gamma_{(3)}\lambda),\nu_2(\bar\lambda,\gamma_{(3)}\lambda)\rangle=0
\end{equation}
in $\widehat{\astring}(\frg)$. Also, using the non-vanishing non-null one-form $v$ and its dual vector field $V$, we can write any $p$-form $\omega_{(p)}\in \Omega^p(\FR^{1,5})$ as $\omega_{(p)}=v\wedge\alpha+*(v\wedge \beta)$. This implies the identity 
\begin{equation}\label{eq:p_form_identity}
\omega_{(p)} = (-1)^{p+1}(\iota_V \omega_{(p)}) \wedge v + *((\iota_V * \omega_{(p)})\wedge v)
\end{equation}
because
\begin{equation}
 (-1)^{p+1}(\iota_V \omega_{(p)}) \wedge v + *((\iota_V * \omega_{(p)})\wedge v)=(-1)^{p+1}\alpha\wedge v+(-1)^{p+1}*(\beta\wedge v)~.
\end{equation}
For $p=6$, the identity reduces to
\begin{equation}\label{eq:6_form_identity}
\omega_{(6)} = -(\iota_V \omega_{(6)})\wedge v~.
\end{equation}
Moreover, a direct computation shows that for any $\omega_{(p)}\in \Omega^3(\FR^{1,5})$,
\begin{equation}\label{eq:3_form_identity}\
\iota_V * \omega_{(p)} = *(\omega_{(p)}\wedge v)~.
\end{equation}

The relevant part of the action, $\CL_{\rm PST}$, is complemented by the terms including $\CH$ from $\CL_{\rm tensor}$. We can combine both into
\begin{equation}
\begin{aligned}
\CL'_{\rm PST} &= \CL_{\rm PST}-\tfrac12\langle\CH,*\CH\rangle+\langle\CH,\nu_2(\bar\lambda,*\gamma_{(3)}\lambda)\rangle\\
&= -\big\langle\iota_{V} (*\CH-\CH-2\nu_2(\bar\lambda,\gamma_{(3)}\lambda)),\CH\big\rangle\wedge v+\big\langle \Phi (\iota_{V} * \CCG), *(\iota_{V}*\CCG)\big\rangle~,
\end{aligned}
\end{equation}
where we also use the fact that $\gamma_{(3)}$ is anti-self-dual. The variation of this expression is readily computed to be
\begin{equation}
\begin{aligned}
\delta \CL'_{\rm PST} &= -2 \big\langle\iota_{V}\CCH,\delta\CH -\tfrac12 \delta v\wedge \iota_{V}\CCH\big\rangle\wedge v+ \langle\CH,\delta\CH\rangle\\
&\phantom{{}={}}  -2 \langle \Phi(\iota_{V} *\CCG),\iota_{V} \CCG\rangle\wedge v\wedge\delta v + 2 \langle \Phi(\iota_{V} *\CCG),\delta \CG\rangle\wedge v\\
&\phantom{{}={}} -2 \langle \Phi(\CCG), \nu_2(\delta\CF,\phi)\rangle-2\langle \Phi(\iota_{V} \CCG),\nu_2(\delta\CF,\phi)\rangle\wedge v\\
&\phantom{{}={}}+\delta_{\phi,\lambda,\chi} \CL'_{\rm PST}~,
\end{aligned}
\end{equation}
where we take into account that $\iota_V \delta v=0$ as $\iota_V v=1$. Inserting the variations of the curvatures given in~\eqref{eq:gen_var_higher_curvs} and using the Bianchi identities~\eqref{eq:bianchi_higher_curvs} to simplify expressions leads to
\begin{equation}
\begin{aligned}
\delta \CL'_{\rm PST} &=2\big\langle \Phi(\iota_{V}*\CCG \wedge v),\mu_1(\Delta D)\big\rangle+\big\langle\mu_1(2\iota_{V}\CCH\wedge v + \CH)-2\nabla \Phi(\iota_{V} *\CCG) \wedge v,\Delta C\big\rangle\\
&\phantom{{}={}}+\big\langle2\dd(\iota_{V}\CCH\wedge v)-\nu_2(\CF,\CF)+\mu_1(\CCG)+4\nu_2(\Phi(\iota_{V}*\CCG\wedge v),\CF),\Delta B\big\rangle\\
&\phantom{{}={}}-\big\langle\nu_2(\CF,2\iota_{V}\CCH\wedge v+\CH)+2\nabla(\nu_2(\Phi(\CCG),\phi)+\nu_2(\Phi(\iota_{V}\CCG \wedge v),\phi))\\
&\phantom{{}={}}+2\nu_2(\Phi(\iota_{V}*\CCG\wedge v),\CH),\delta A\big\rangle-(\big\langle\iota_{V}\CCH,\iota_{V}\CCH\big\rangle+2 \big\langle \Phi(\iota_{V} *\CCG),\iota_{V} \CCG\big\rangle)\wedge v\wedge \delta v\\
&\phantom{{}={}}+\delta_{\phi,\lambda,\chi} \CL'_{\rm PST}~.
\end{aligned}
\end{equation}
Note that we can cancel some of the terms originating from $\langle\CH,\delta\CH\rangle$ by adding the variation of the topological term, which is given by
\begin{equation}
\delta \CL_{\rm top} = \big\langle\nu_2(\CF,\CF) + \mu_1(\CG),\Delta B\big\rangle -\big\langle\mu_1(\CH),\Delta C\big\rangle -\big\langle\nu_2(\CF,\CH),\delta A\big\rangle~.
\end{equation}
Furthermore, there are additional terms for the variation with respect to the gauge potential $A$ coming from both $\CL_{\rm tensor}$ and $\CL_{\rm hyper}$. After including these terms, and again using the Bianchi identities~\eqref{eq:bianchi_higher_curvs} to simplify expressions, we arrive at
\begin{equation}\label{eq:lagrangian_part_variation}
\begin{aligned}
\delta\CL &=2\big\langle \Phi(\iota_{V}*\CCG \wedge v),\mu_1(\Delta D)\big\rangle+\big\langle 2\mu_1(\iota_{V}\CCH\wedge v)-2\nabla \Phi(\iota_{V} *\CCG) \wedge v,\Delta C\big\rangle\\
&\phantom{{}={}}+\big\langle2\dd(\iota_{V}\CCH\wedge v)+2\mu_1(\CCG)+4\nu_2(\Phi(\iota_{V}*\CCG\wedge v),\CF),\Delta B\big\rangle\\
&\phantom{{}={}}+\big\langle\mu_1(\CCI)- 2\nu_2(\CF,\iota_{V}\CCH\wedge v)+2\nabla(\nu_2(\Phi(\iota_{V}\CCG \wedge v),\phi))\\
&\phantom{{}={}}-2\nu_2(\Phi(\iota_{V}*\CCG\wedge v),\CH),\delta A\big\rangle-\big(\big\langle\iota_{V}\CCH,\iota_{V}\CCH\big\rangle+2 \big\langle \Phi(\iota_{V} *\CCG),\iota_{V} \CCG\big\rangle\big)\wedge v\wedge \delta v\\
&\phantom{{}={}}+\delta_{\phi,\chi,\lambda,Y,q,\psi} \CL~,
\end{aligned}
\end{equation}
where we also used $\nabla(\CCG+\nu_2(\Phi(\CCG),\phi))=\dd^2C=0$. Given this, it is immediate that the Lagrangian is invariant under any one of the symmetry transformations
\begin{equation}\label{eq:simple_lagrangian_symmetries}
\delta A = \varphi_A\wedge v~,~~~\Delta B = \varphi_B \wedge v~,~~~\Delta C = \varphi_C \wedge v~,~~~\Delta D = \varphi_D \wedge v~,
\end{equation}
where $\varphi_C$ and $\varphi_D$ are free parameters taking values in $\frg^*[2]\oplus \FR[2]$ and $\frg^*[3]$, respectively, while $\varphi_A$ is restricted to lie in $\FR^*$ and $\varphi_B$ is restricted to lie in $\FR[1]$. Furthermore, it can be shown using the above variation that the Lagrangian is invariant under the combined transformations
\begin{gather}
\nonumber\delta v=  \dd \varphi_v(x)~,~~~\delta A = \varphi_v(x)\Phi(\iota_V *\CCG)~,\\
\Delta B = \varphi_v(x) \iota_V\CCH~,~~~\Delta C = -\varphi_v(x) \iota_V \CCG~,\\
\nonumber\Delta D= \tfrac12 \varphi_v(x)\iota_V(\CCI)~,
\end{gather}
where $\varphi_v$ is a function on $\FR^{1,5}$. This symmetry transformation exposes the auxiliary nature of $v$, guaranteeing that no additional degrees of freedom are introduced.

Let us now come to the derivation of the duality equations from the variation~\eqref{eq:lagrangian_part_variation}.
Starting with the variation with respect to $\mu_1(\Delta D)$ we have
\begin{equation}
\Phi(\iota_{V} *\CCG)\wedge v{\big|}_{\frg} = 0~.
\end{equation}
Since, by construction, $\iota_{V}*\CCG$ has no common directions with $v$ and, furthermore, the kernel of $\Phi$ lies in $\FR[2]$, this is equivalent to
\begin{equation}\label{eq:proj1_G1}
\iota_{V}*\CCG{\big|}_{\frg^*[2]} = 0~.
\end{equation}
Additionally, we can use the last symmetry in~\eqref{eq:simple_lagrangian_symmetries} to gauge away $\iota_V \CCG{\big|}_{\frg^*[2]}$. Indeed, from~\eqref{eq:gen_var_higher_curvs} we have,
\begin{equation}
\iota_V \delta_D\CCG \wedge v = \iota_V \mu_1(\Delta D) \wedge v = -\mu_1(\varphi_D)\wedge v~.
\end{equation}
Thus, choosing $\mu_1(\varphi_D) = -\iota_V \CCG$ we gauge-fix $\iota_V \CCG{\big|}_{\frg^*[2]}=0$, which in conjunction with equation~\eqref{eq:proj1_G1} implies
\begin{equation}\label{eq:proj1_CCG}
\CCG{\big|}_{\frg^*[2]}= 0~.
\end{equation}
This reduces the variation with respect to $\Delta C$ to the equation
\begin{equation}
\mu_1(\iota_V\CCH\wedge v)=0~.
\end{equation}
As, again by construction, $\iota_V\CCH$ does not share directions with $v$, we can write this as
\begin{equation}\label{eq:proj1_CCH}
\iota_V \CCH{\big|}_{\FR^*[1]}=0~.
\end{equation}
Taking~\eqref{eq:proj1_CCG} into account and turning our attention to the variation with respect to $\Delta B$ we have
\begin{equation}
\dd(\iota_V\CCH\wedge v)+\mu_1(\CCG) =0~.
\end{equation}
This immediately implies $\mu_1(\CCG)\wedge v =0$ which is equivalent to 
\begin{equation}\label{eq:proj2_CCG}
\iota_V * \CCG{\big|}_{\FR[2]} = 0~.
\end{equation}
Additionally, using the third symmetry in~\eqref{eq:simple_lagrangian_symmetries} we have
\begin{equation}
\iota_V\delta_C\CCG\wedge v {\big|}_{\FR[2]} = \iota_V \dd\Delta C\wedge v{\big|}_{\FR[2]} = -\dd\varphi_C\wedge v{\big|}_{\FR[2]}~, 
\end{equation}
which when choosing $\varphi_C = -\iota_V\CCG{\big|}_{\FR[2]}$ allows to gauge-fix to $\iota_V\CCG{\big|}_{\FR[2]}=0$. This, together with~\eqref{eq:proj1_CCG} and~\eqref{eq:proj2_CCG}, leads to the first duality equation $\CCG = 0~.$\\
Using this leaves the variation with respect to $\Delta B$ with the equation
\begin{equation}
\dd (\iota_V \CCH \wedge v){\big|}_{\FR[1]} = 0~,
\end{equation}
which has the general solution
\begin{equation}
\iota_V \CCH\wedge v{\big|}_{\FR[1]} = \dd \tilde\varphi\wedge v~,
\end{equation}
where $\tilde \varphi \in \Omega^1(\FR^{1,5})\otimes\FR[1]~.$ Note, that using~\eqref{eq:gen_var_higher_curvs} we also have under the first symmetry in~\eqref{eq:simple_lagrangian_symmetries} that
\begin{equation}
\iota_V\delta\CCH = \iota_V(*(\dd\varphi_B\wedge v)-\dd\varphi_B\wedge v)\wedge v = -\dd \varphi_B\wedge v~,
\end{equation}
where $\varphi_B$ is also an element of $\Omega^1(\FR^{1,5})\otimes\FR[1]~.$ As this has the same form as the general solution above, we can gauge-fix to $\varphi_B = - \tilde\varphi$ and arrive at the self-duality equation $\CCH = 0$. Lastly, looking at the variation with respect to $\delta A$ and taking into account all equations of motion we have derived so far, we are left with the last duality equation $\CCI = 0$. Note that this leaves the equations coming from the variation with respect to $\delta v$ trivially satisfied.

The remaining equations of motion are straightforward to calculate and, altogether, we arrive at the set of equations
\begin{equation}
\begin{aligned}
\CCH&=*\CH - \CH -\nu_2(\bar\lambda,*\gamma_{(3)}\lambda)=0 ~,\\
\CCG&=\CG -\nu_2(*\CF,\phi)+2\nu_2(\bar\lambda,*\gamma_{(2)}\chi)=0~,\\
\CCI&=\CI +\mu_2(\nu_2(\bar\lambda,\phi),\gamma^\mu\lambda)\vol_\mu+2\langlec-\acton q,*\nabla q\ranglec-2\langlec\bar\psi,-\acton\gamma^\mu\psi\ranglec\vol_\mu=0~,
\end{aligned}
\end{equation}
together with the remaining equations of motion for the tensor multiplet,
\begin{equation}
\begin{aligned}
\nu_2(\slasha{\nabla}\lambda_i,\phi)+\tfrac12\nu_2(\lambda_i,\slasha{\dd}\phi) &= -\tfrac12*\nu_2(\CF,*\gamma_{(2)}\chi_i) - \nu_2(Y_{ij},\chi^j)+\tfrac18*\nu_2(*\gamma_{(3)}\lambda_i,\CH)\\
&\phantom{{}={}}+\mu_1(\langlec-\acton q_i,\psi\ranglec)~,\\
\nu_2(Y^{ij},\phi)-2\nu_2(\bar\lambda^{(i},\chi^{j)}) &= -\tfrac12 \mu_1(\langlec q^{(i},-\acton q^{j)}\ranglec)~,\\
\slasha{\dd} \chi_i &= \tfrac12*\nu_2(\CF,*\gamma_{(2)}\lambda_i)+2\nu_2(Y_{ji},\lambda^j)~,\\
\square \phi &= -*\nu_2(\CF,*\CF) - \nu_2(Y_{ij},Y^{ij}) +\nu_2(\bar\lambda,\slasha{\nabla}\lambda)~,
\end{aligned}
\end{equation}
as well as the equations for the hypermultiplet,
\begin{equation}
\begin{aligned}
\square q_i &= -4\bar\lambda_i\acton\psi -2 Y_{ij}\acton q^j ~,\\
\slasha{\nabla} \psi &= 2 \lambda \acton q~.
\end{aligned}
\end{equation}
Note that these equations of motion become partially degenerate for $\phi_s=0$; in particular, the duality equation linking $\CG$ and $\CF$ breaks down. This is again a reflection of the tensionless string phase transition~\cite{Seiberg:1996vs,Duff:1996cf}, see also~\cite{Samtleben:2012fb}.

\subsection{BPS states}\label{ssec:BPS_states}

Recall that abelian self-dual string solitons are described by a 2-form $B\in\Omega^2(\FR^4)\otimes \au(1)$ and a Higgs field $\phi\in\Omega^0(\FR^4)\otimes \au(1)$ which satisfy the equation
\begin{equation}\label{eq:abelian_SDS}
 H:=\dd B=*\dd \phi~.
\end{equation}
They form BPS states of the (2,0)-theory~\cite{Howe:1997ue}. An appropriate non-abelian generalization of~\eqref{eq:abelian_SDS} using string structures was derived in~\cite{Saemann:2017rjm}. The resulting equations are
\begin{equation}\label{eq:eom_sds}
 H:=\dd B-(A,\dd A)-\tfrac{1}{3}(A,[A,A])=*\dd \phi~,~~~F_L=*F_L~,~~~F_R=-*F_R~,
\end{equation}
where $A\in \Omega^1(\FR^4)\otimes (\frg_L\oplus \frg_R)$ is a Lie-algebra valued one-form with curvature
\begin{equation}
 F=F_L+F_R~,~~~F_L=\dd A_L+\tfrac12[A_L,A_L]~,~~~F_R=\dd A_R+\tfrac12[A_R,A_R]~.
\end{equation}
For $A,B$ to be a connection on a global string structure, we need to demand that the first Pontrjagin class $\tfrac12p_1(F)$ of $F$  vanishes, which is the case if
\begin{equation}
 \tfrac12p_1(F_L)+\tfrac12p_1(F_R)=0~.
\end{equation}
Note that closely related constructions have been discussed in the past, see e.g.~\cite{Duff:1996cf} or~\cite{Akyol:2012cq}. 

Equations~\eqref{eq:eom_sds} passes many consistency checks and fulfills expectations that one would have from a non-abelian generalization of the self-dual string soliton~\cite{Saemann:2017rjm}. First, it is indeed a well-motivated analogue of the 't~Hooft--Polyakov monopole on $\FR^3$ for $\frg=\aspin(3)\cong \asu(2)$. The string Lie 2-algebra $\astring(\asu(2))$ is a categorified analogue of $\asu(2)=\aspin(3)$ and the equations~\eqref{eq:eom_sds} dimensionally reduce to the non-abelian Bogomolny monopole equations. Topological considerations also work as expected. Finally, these equations are formulated in the language of the string 2-group model $\sString_{\rm sk}(n)$. As shown in~\cite{Saemann:2017rjm}, they can be rephrased in the string 2-group model $\sString_{\hat\Omega}(n)$ such that gauge orbits and gauge transformations are mapped consistently into each other. 

Considering now the supersymmetry transformations~\eqref{eq:susy_trafo_tensor} and~\eqref{eq:susy_trafo_vector}, it is clear that equations~\eqref{eq:eom_sds} are indeed the equations for BPS states constant along the temporal and one spatial direction, see~\cite{Akyol:2012cq} for a closely related observation. Solutions to~\eqref{eq:eom_sds} also satisfy the equations of motion of our model. For this to work, however, we have to regard $\phi$ as the scalar field arising from $B$ in the constant directions: $\phi=B_{05}$.

If we now believe that our action can be quantized (which would require the full machinery of BV quantization), we might speculate that it is useful to include self-dual string operators into the path integral as categorified analogues of monopole operators. In particular, recall that monopole operators were shown to enhance the $\CN=6$ supersymmetry of the ABJM model to the full, expected $\CN=8$ supersymmetry, see~\cite{Bashkirov:2010kz} and references therein. Something similar might happen in our model, but it remains unclear how the phase transition at $\phi_s=0$ should be addressed.

\subsection{Formulation for other string 2-group models}

In the above, we have used the skeletal model $\astring_{\rm sk}(\frg)$ as the basis of our gauge structure as this readily translates to the $(1,0)$-gauge structures of~\cite{Samtleben:2011fj}. However, as described in section~\ref{ssec:string_Lie_2}, there are multiple categorically equivalent models for the string 2-group and one would expect that our Lagrangian admits equivalent formulations based on these other string 2-group models.

In particular, there should be an equivalent formulation for the string 2-group model based on path and loop spaces of~\cite{Baez:2005sn}. The corresponding string Lie 2-algebra is given by
\begin{equation}
\astring_{\hat\Omega}(\frg) =\ \big(~ P_0\frg~\xleftarrow{~\mu_1~}~\Omega\frg\oplus\FR~\big)~,
\end{equation}
where $P_0\frg$ and $\Omega\frg$ are based path and based loop spaces of the vector space $\frg$, respectively, and $\mu_1$ is the embedding of $\Omega\frg$ into $P_0\frg$. The obvious Lie brackets give rise to the following non-trivial higher products:
\begin{equation}
 \begin{aligned}
  \mu_1&:\Omega\frg\oplus\FR\rightarrow P_0\frg~,~~~\hspace{2.4cm}\mu_1((\lambda,r))=\lambda~,\\
  \mu_2&:P_0\frg\wedge P_0\frg\rightarrow P_0\frg~,~~~\hspace{2cm}\mu_2(\gamma_1,\gamma_2)=[\gamma_1,\gamma_2]~,\\
  \mu_2&:P_0\frg\otimes(\Omega\frg\oplus\FR)\rightarrow \Omega\frg\oplus\FR~,~~~\\
  &\hspace{4cm}\mu_2\big(\gamma,(\lambda,r)\big)=\left([\gamma,\lambda]\; ,\; -2\int_0^1 \dd\tau \left(\gamma(\tau),\dder{\tau}\lambda(\tau)\right)\right)~.
 \end{aligned}
\end{equation}

To define string structures using this string Lie 2-algebra model, one needs to introduce the additional maps~\cite{Saemann:2017rjm}
\begin{equation}
\begin{aligned}
 \nu_2:P_0\frg\times P_0\frg&\rightarrow \Omega\frg\oplus \FR~,\\
 \nu_2(\gamma_1,\gamma_2)&\coloneqq \left(\chi([\gamma_1,\gamma_2])\;,\;2\int_0^1\dd \tau (\dot \gamma_1,\gamma_2)\right)~,
\end{aligned}
\end{equation}
where $\chi: P_0\frg \rightarrow \Omega\frg\oplus\FR$ is given by $\chi(\gamma) = \lambda-f\cdot\partial\lambda$ for an arbitrarily chosen, smooth function $f:[0,1]\rightarrow \FR$ with $f(0)=0$ and $f(1)=1$. With this the appropriate expressions for the two- and three-form curvatures are then given by
\begin{equation}\label{eq:loop_curvatures}
\begin{aligned}
\CF &= \dd A + \tfrac12 \mu_2(A,A) + \mu_1(B)~,\\
\CH &= \dd B + \mu_2(A,B) - \nu_2(A,\CF)~,
\end{aligned}
\end{equation}
where the term $-\nu_2(A,\CF)$ can be regarded as the counterpart to the Chern--Simons term $\nu_2(A,\dd A)+\tfrac13\nu_2(A,\mu_2(A,A))$ in the expression for $\CH$ in the skeletal case. 

Analogously to the skeletal case we can now minimally extend this Lie 2-algebra to a cyclic Lie 3-algebra, which leads us to the complex
\begin{equation}
\xymatrixcolsep{4pc}
\xymatrixrowsep{1pc}
\myxymatrix{
 \FR^*\ar@{}[d]|{\oplus}\ar@{<-}[r]^{\mu_1=(0,\id)^*} & \hat\Omega\frg^*[1]\ar@{}[d]|{\oplus}\ar@{<-}[r]^{\mu_1=(\id,0)^*} & P_0\frg^*[2]\ar@{}[d]|{\oplus}\ar@{<-^{)}}[r]^{\mu_1=\partial^*} & ~\frg^*[3]\\
 P_0\frg\ar@{<-}[r]^{\mu_1=(\id,0)}& \hat\Omega\frg[1]\ar@{<-}[r]^{\mu_1=(0,\id)} & \FR[2] & }
\end{equation}
where $\dpar:P_0\frg\rightarrow \frg$ is the endpoint evaluation map, $\dpar^*$ its dual, etc.

However, given the curvatures as in~\eqref{eq:loop_curvatures} above, this picture cannot be realized as a
$(1,0)$-gauge structure of~\cite{Samtleben:2011fj}. In particular, the $(1,0)$-gauge structure requires the map $\nu_2$ to be symmetric and does not allow an anti-symmetric term of the form $\mu_2(A,B)$. Both of these requirments are not satisfied for $\astring_{\hat\Omega}(\frg)$. This indicates that the $(1,0)$-gauge structure is too strict and does not encapsulate the full picture. A full formulation of our model in this picture should therefore shed more light on the situation and we plan to develop this in a future publication.

\section{Dimensional reduction}\label{sec:dim_reductions}

\subsection{Reduction to super Yang--Mills theory}\label{ssec:4d_reduction}

Let us begin with a first consistency check: the dimensional reduction of our model to a supersymmetric Yang--Mills theory. This is an extension of the discussion in~\cite{Saemann:2017rjm}, where the $(1,0)$-tensor multiplet part of the action~\eqref{eq:full_action} was considered.

We start from the Lie 3-algebra $\widehat{\astring}(\asu(n))$. We then compactify $\FR^{1,5}$ along two spatial directions on a torus $T^2$ with radii $R_9$ and $R_{10}$ and modular parameter 
\begin{equation}
 \tau=\tau_1+\di \tau_2=\frac{\theta}{2\pi}+\frac{\di}{g^2_{\rm YM}}~,
\end{equation}
where we already indicated the expected identification with coupling constants in four dimensions. We assume that, analogously to the case of M2-brane models~\cite{Mukhi:2008ux}, the component $\phi_s\in\FR^*[1]$ of $\phi$ develops a vacuum expectation
\begin{equation}
 \langle \phi_s \rangle=-\frac{1}{32\pi^2} \frac{1}{R_{10}^2}=-\frac{1}{32\pi^2}\frac{\tau_2}{R_9 R_{10}}~,
\end{equation}
which matches the inverse length dimension~2 of the scalar field $\phi$. In order to get the full $\theta$-term, we also put a constant 2-form field on the torus:
\begin{equation}
 \langle B_s \rangle=\frac{1}{16\pi^2}\frac{\tau_1}{R_9 R_{10}} {\rm vol}(T^2)~.
\end{equation}
We are then interested in the double scaling limit of small radii $R_9$ and $R_{10}$ with the ratio $R_9/R_{10}$ constant. Note that 
\begin{equation}
 -2\int_{T^2} {\rm vol}~(T^2)\langle\phi_s\rangle=\frac{1}{4g_s}=\frac{1}{4g^2_{\rm YM}}\eand \int_{T^2}\langle B_s\rangle=\frac{\tau_1}{4}=\frac{\theta}{8\pi}~.
\end{equation}
according to the usual relation between compactification radii, string coupling constant and Yang--Mills coupling in four dimensions.

We can regard $\phi_s$ as the radial coordinate on a cone over the target space $\FR^{2\times 2n}$, and scaling $\phi_s$ involves a dilation of the hyper Kähler cone $\FR^{2\times 2n}$. Considering the underlying geometric structures as presented e.g.~in~\cite{Samtleben:2012fb}, we note that the homothetic Killing spinor rescales the metric on $\FR^{2\times 2n}$, which is readily identified with a rescaling of the symplectic form $\Omega$ defining the bilinear pairing $\langlec-,-\ranglec$, again by a factor of $\frac{1}{\pi^2R_9R_{10}}$. In the small radius limit, the dominant terms in the Lagrangian are therefore
\begin{equation}
 \begin{aligned}
  \CL_{R\rightarrow 0}&= \frac{1}{\pi^2R_9R_{10}}\Big[\frac{\tau_1}{4}(F,F)+\frac{\tau_2}{4}\Big((F,*F)-2\vol(Y_{ij},Y^{ij})+4\vol(\bar\lambda,\slasha{\nabla} \lambda)\\
  &-\langlec \nabla q,*\nabla q\ranglec + 2\vol\langlec\bar\psi,\slasha{\nabla}\psi\ranglec + 8\vol\langlec\bar\psi, \lambda_i\acton q^i\ranglec + 2\vol\,\langlec q^i,Y_{ij}\acton q^j\ranglec\Big)\Big]~,
 \end{aligned}
\end{equation}
where we used the fact that $\nu_2(\CF,*\CF)|_{\FR[1]}=(F,*F)$ and $\nu_2(\CF,\CF)|_{\FR[1]}=(F,F)$  in the case at hand. We can now reduce the six-dimensional gauge potential $A$ to a four dimensional one, $\check A$, with curvature $\check F$ together with two scalar fields $\check \sigma$, which are the components of $A$ along the torus. We also rotate the field content in the hypermultiplet to obtain scalar fields $q^i$ and spinors $\psi$ taking values in the adjoint representation of $\asu(n)$. Finally we integrate out the auxiliary field $Y$ and integrate over the torus to implement the compactification. This yields the Lagrangian
\begin{equation}
 \begin{aligned}
    \CL_{4d}= \frac{1}{4g^2_{\rm YM}}\Big(&(\check F,*\check F)+\tr(\check \nabla \check \sigma,*\check \nabla \check \sigma)+4\vol(\bar\lambda,\check{\slasha{\nabla}} \lambda)+\tr(\check \nabla q,*\check \nabla q) + 4\vol\tr(\bar\psi,\check{\slasha{\nabla}}\psi)\\
    &+ 4\vol\tr(\bar\lambda, [\slasha{\check{\sigma}},\lambda])+ 4\vol\tr(\bar\psi, [\slasha{\check{\sigma}},\psi])+ 8\vol\tr(\bar\psi, [\lambda,q])\\
    &+\tr([\check \sigma_1,\check \sigma_2]^2)+\tr([\check \sigma,q]^2)+ \tr([q^1,q^2]^2)\Big)+\frac{\theta}{8\pi}(F,F)~,
 \end{aligned}
\end{equation}
which is a supersymmetric gauge theory in four dimensions with an $\CN=2$ vector multiplet coupled to an $\CN=2$ hypermultiplet and has underlying gauge Lie algebra $\asu(n)$.

Analogous reductions are clearly possible for the Lie 3-algebras $\widehat{\astring}(\frg)$ with $\frg$ any other Lie algebra of type $D$ or $E$. The four-dimensional theory will then have gauge Lie algebra $\frg$.

It is not too surprising that we are able to reduce our model to super Yang--Mills theory in four dimensions because it contains a free vector multiplet in six dimensions. The fact, however, that this reduction is compatible with the supersymmetry mixing the vector and tensor multiplets and that it reproduces the expected $\theta$-term is very satisfying.

\subsection{Reduction to supersymmetric Chern--Simons-matter theories}\label{ssec:3d_reduction}

There is no direct argument within M-theory that an effective description of M5-branes should be reducible to one of M2-branes. However, one would expect that some form of T-duality which allows for analogous reductions in string theory should still exist, cf.~e.g.~\cite{Lambert:2016xbs}. Moreover, the fact that M2-branes can end on M5-branes has led to attempts of constructing M5-brane models from the M2-brane models, see e.g.~\cite{Ho:2008nn,Ho:2008ve}, which again suggests a link between M5-brane and M2-brane models. Finally, note that while the M2-brane models seem very different from the M5-brane models, the former can be recast in the form of a higher gauge theory~\cite{Palmer:2013ena}.

We start from our model~\eqref{eq:full_action} for Lie 3-algebra $\widehat{\astring}(\au(n)\times \au(n))$, but in particular the case $\widehat{\astring}(\asu(2)\times \asu(2))$ should be interesting for further investigations. We choose a metric of split signature on $\au(n)\times \au(n)$, anticipating this to become the gauge Lie algebra of the M2-brane model. We then compactify $\FR^{1,5}$ to $\FR^{1,2}\times S^3$, but a more general choice of compact 3-dimensional spin manifold $M^3$ than $S^3$ should also suffice. 

The general dimensional reduction will yield a rather general deformation of the ABJM model. For simplicity, we shall restrict the fields rather severely. While this reduces the supersymmetry of the model, it makes the interpretation of the resulting action clearer. We decompose the fields $(B,\phi,\chi)$ in the tensor multiplets taking values in $\FR[1]\oplus \FR^*[1]$ as $\phi=\phi_r+\phi_s$, etc. We then restrict to
\begin{equation}
 B_r=0~,~~~\phi_r=0~,~~~\chi_r=0~.
\end{equation}
Also, $B_s$ is the potential for a gerbe over $S^3$ with Dixmier--Douady class $k$ such that
\begin{equation}
 \int_{S^3} \dd B_s=\frac{k}{2\pi}
\end{equation}
and $B_s$ has no further components. We also restrict the gauge potential such that its components $A_{3,4,5}$ along $S^3$ vanish. Correspondingly, we demand that the spinors satisfy $\iota_{\der{x^{0,1,2}}}*\lambdab \gamma_{(3)}\lambda=0$. With these constraints, the kinematical term for the 3-form curvature reduces according to
\begin{equation}
\begin{aligned}
 -\tfrac12 \int_{S^3} \langle \CH,*\CH\rangle&=-\tfrac12 \int_{S^3} 2\langle \dd B_r,-(A,\dd A)-\tfrac13(A,[A,A])\rangle\\
 &=\frac{k}{2\pi}(A,\dd A)+\tfrac13(A,[A,A])~.
\end{aligned}
\end{equation}
We thus obtain the Lagrangian for Chern--Simons theory, and the quantized coupling constant arises from the topological class describing the gerbe over the compactifying 3-manifold $M^3$.

Let us also consider the PST terms in the action~\eqref{eq:full_action}. It makes sense to restrict the non-vanishing vector field $V$ to be a section of the tangent bundle of $\FR^{1,2}$. Then 
\begin{equation}
\begin{aligned}
\tfrac12\big\langle\iota_{V}\CCH,\CCH\big\rangle\wedge v&=\frac{k}{\pi}{\rm cs}(A)-\frac{k}{\pi}(\bar\lambda,*\gamma_{345}\lambda)~,\\
\end{aligned}
\end{equation}
and we merely get a further contribution to the Chern--Simons term. Altogether, we obtain a supersymmetric Chern--Simons-matter theory coupled to an additional Yang--Mills component with coupling constant $\phi_s$, just as in the last section. Again, it makes sense to set $\phi_s=\frac{1}{4g_{\rm YM}^2}$ to obtain an interacting Chern--Simons-matter theory.

\subsection{Comment on a reduction to an \texorpdfstring{$L_\infty$}{Linfinity}-algebra model}\label{ssec:0d_reduction}

Finally, let us briefly comment on a full dimensional reduction of the action~\eqref{eq:full_action}. The resulting model is a categorified matrix model or {\em $L_\infty$-algebra model} of the type studied in~\cite{Ritter:2013wpa,Ritter:2015ymv}. It may be regarded as a deformed M-theory analogue to the IKKT model~\cite{Ishibashi:1996xs}. In particular, it is rather obvious that the mechanism described in section~\ref{ssec:4d_reduction} can be applied to the fully reduced model to obtain a deformation of the IKKT model.

The resulting $L_\infty$-algebra model is very interesting as it should be able to shed some light on the so far rather opaque problem of higher quantization, cf.~\cite{Bunk:2016rta}. This quantization procedure starts from a multisymplectic phase space where the multisymplectic form describes the topological class of a categorified prequantum line bundle. Many aspects of this procedure remain poorly understood. In particular, it is unclear how a polarization should be imposed on sections of categorified line bundles and for spaces with multisymplectic forms which are not torsion, even the unpolarized sections are hard to control. Nevertheless, quantized multisymplectic spaces appear within M-theory. For example, lifting the fuzzy funnel of D1-branes ending on D3-branes to M-theory suggest that the worldvolume of M2-branes ending on M5-branes polarizes into 3-spheres quantized as multisymplectic phase spaces with the multisymplectic form being the volume form.

Recall that the IKKT model has interesting noncommutative spaces as stable classical solutions. Correspondingly, we would expect a suitable $L_\infty$-algebra model to allow for interesting higher quantized spaces as classical solutions. This would circumvent the construction of a categorified Hilbert space in higher quantization and jump directly to the higher analogue of an operator algebra. We shall leave the exploration of these reductions to future publications.

\section*{Acknowledgements}

We would like to thank David Berman, Leron Borsten, Werner Nahm, Denjoe O'Connor, Daniel Thompson and Martin Wolf for useful conversations. We are also very grateful to contributors to the useful discussion of a first version of this paper on The $n$-Category Caf\'e, especially John Baez, David Ben--Zvi, Jacques Distler and Sammuel Monnier. We are particularly indebted to Urs Schreiber and his large body of work developing and explaining higher structures. The work of L.S.~was supported by an STFC PhD studentship.

\bibliography{bigone}

\bibliographystyle{latexeu}

\end{document}